\documentclass[twocolumn]{aastex63}

\def\cii{\hbox{{\rm [C {\scriptsize II}]}}}

\def\13cii{\hbox{{\rm [$^{13}$C {\scriptsize II}]}}}

\def\ci{\hbox{{\rm [C {\scriptsize I}]}}}
\def\oi{\hbox{{\rm [O {\scriptsize I}]}}}
\def\hii{\hbox{{\rm H {\scriptsize II}}}}
\def\siv{\hbox{{\rm S {\scriptsize IV}}}}
\def\neii{\hbox{{\rm Ne {\scriptsize II}}}}
\def\neiii{\hbox{{\rm Ne {\scriptsize III}}}}
\def\siii{\hbox{{\rm S {\scriptsize III}}}}

\defcitealias{MSP1991}{MSP}
\defcitealias{TFT2007}{TFT07}
\defcitealias{VA2013}{VA13}
\defcitealias{VPHAS_Wd2_2015}{VPHAS+}

\usepackage{gensymb}
\usepackage[flushleft]{threeparttable}
\usepackage{amsmath}
\usepackage{changepage}
\usepackage{rotating}
\usepackage{array}
\usepackage{multirow}

\received{March 25, 2022}
\accepted{August 16, 2022}

\shorttitle{RCW~49}
\shortauthors{Tiwari et al.}

\graphicspath{{./}{figures/}}

\begin{document}

\title{SOFIA FEEDBACK survey: PDR diagnostics of stellar feedback in different regions of RCW~49}


\author{M. Tiwari}
\affiliation{University of Maryland, Department of Astronomy, College Park, MD 20742-2421, USA}
\affiliation{Max-Planck Institute for Radioastronomy, Auf dem H\"{u}gel, 53121 Bonn, Germany}

\author{M. Wolfire}
\affiliation{University of Maryland, Department of Astronomy, College Park, MD 20742-2421, USA}
\author{M. W. Pound}
\affiliation{University of Maryland, Department of Astronomy, College Park, MD 20742-2421, USA}
\author{E. Tarantino}
\affiliation{University of Maryland, Department of Astronomy, College Park, MD 20742-2421, USA}
\author{R. Karim}
\affiliation{University of Maryland, Department of Astronomy, College Park, MD 20742-2421, USA}


\author{L. Bonne}
\affiliation{SOFIA Science Center, USRA, NASA Ames Research Center, M.S. N232-12, Moffett Field, CA 94035, USA}

\author{C. Buchbender}
\affiliation{I. Physik. Institut, University of Cologne, Z\"{u}lpicher Str. 77, 50937 Cologne, Germany}

\author{ R. G\"{u}sten}
\affiliation{Max-Planck Institute for Radioastronomy, Auf dem H\"{u}gel, 53121 Bonn, Germany}

\author{C. Guevara}
\affiliation{I. Physik. Institut, University of Cologne, Z\"{u}lpicher Str. 77, 50937 Cologne, Germany}


\author{S. Kabanovic}
\affiliation{I. Physik. Institut, University of Cologne, Z\"{u}lpicher Str. 77, 50937 Cologne, Germany}

\author{\"{U}. Kavak}
\affiliation{SOFIA Science Center, USRA, NASA Ames Research Center, M.S. N232-12, Moffett Field, CA 94035, USA}
\author{M. Mertens}
\affiliation{I. Physik. Institut, University of Cologne, Z\"{u}lpicher Str. 77, 50937 Cologne, Germany}

\author{N. Schneider}
\affiliation{I. Physik. Institut, University of Cologne, Z\"{u}lpicher Str. 77, 50937 Cologne, Germany}

\author{R. Simon}
\affiliation{I. Physik. Institut, University of Cologne, Z\"{u}lpicher Str. 77, 50937 Cologne, Germany}

\author{J. Stutzki}
\affiliation{I. Physik. Institut, University of Cologne, Z\"{u}lpicher Str. 77, 50937 Cologne, Germany}

\author{A. G. G. M. Tielens}
\affiliation{University of Maryland, Department of Astronomy, College Park, MD 20742-2421, USA}
\affiliation{Leiden Observatory, Leiden University, PO Box 9513, 2300 RA Leiden, Netherlands.}

\begin{abstract}
We quantified the effects of stellar feedback in RCW~49 by determining the physical conditions in different regions using the \cii\,~158~$\mu$m and \oi\,~63~$\mu$m observations from SOFIA, the $^{12}$CO~(3-2) observations from APEX and the H$_2$ line observations from Spitzer telescopes. Large maps of RCW~49 were observed with the SOFIA and APEX telescopes, while the Spitzer observations were only available towards three small areas. 
From our qualitative analysis, we found that the H$_2$~0-0~S(2) emission line probes denser gas compared to the H$_2$~0-0~S(1) line. In four regions (``northern cloud'', ``pillar'', ''ridge'', and ``shell''), we compared our observations with the updated PDR Toolbox models and derived the integrated far-ultraviolet flux between 6--13.6~eV ($G_{\rm 0}$), H nucleus density ($n$), temperatures and pressures. 
We found the ridge to have the highest $G_{\rm 0}$ (2.4 $\times$ 10$^3$~Habing~units), while the northern cloud has the lowest $G_{\rm 0}$ (5 $\times$ 10$^2$~Habing~units). This is a direct consequence of the location of these regions with respect to the Wd2 cluster. The ridge also has a high density (6.4 $\times$ 10$^3$~cm$^{-3}$), which is consistent with its ongoing star formation. 
Among the Spitzer positions, we found the one closest to the Wd2 cluster to be the densest, suggesting an early phase of star formation. Furthermore, the Spitzer position that overlaps with the shell was found to have the highest $G_{\rm 0}$ and we expect this to be a result of its proximity to an O9V star.
 
\end{abstract}

\keywords{ISM --- Photodissociation regions --- Molecular clouds}

\section{Introduction} \label{sec:intro}
Stellar feedback is one of the most important ingredients in the evolution of the interstellar medium (ISM). In particular, massive stars play a key role in regulating different feedback mechanisms (for recent observational studies see \citealt{Pabst2019}, \citealt{Luisi2021}, \citealt{Tiwari2021} and \citealt{Kavak2022}). They inject immense amounts of energy into their surroundings through stellar winds and through extreme ultra-violet (EUV, $h\nu$ $>$ 13.6 eV) and far-UV (FUV, 6 $<$ $h\nu$ $<$ 13.6~eV) photons. The EUV photons ionise gas (H $\to$ H$^+$) in the vicinity of a star giving rise to \hii\ regions. The FUV photons heat the gas via the photoelectric effect on small grains and PAHs, dissociate molecules, and ionise atoms e.g., C $\to$ C$^+$, giving rise to photo-dissociation regions (PDRs). This mechanical and radiative energy input catalyses various physical and chemical processes, which shape the dense and diffuse gas into molecular clouds and into structures like shells and pillars. These structures vary in their physical conditions based on their location with respect to the illuminating source and on the geometry of the surrounding medium.  

Determination of physical conditions in these molecular clouds, shells and pillars is done through various chemical species. Neutral atomic gas in the ISM cools mainly at far-infrared wavelengths
through \cii\ (158~$\mu$m) and \oi\ (63 and 145~$\mu$m) line emission and probes regions
predominantly at  $A_{\rm v}$ $<$ 3. Molecular gas cooling is dominated by $^{12}$CO rotational emission which probes regions at larger $A_{\rm v}$ (for details see \citealt{Hollenbach1999,Wolfire2003,Wolfire2022}).\\

RCW~49 is one of the most luminous massive star-forming regions in our Galaxy. It is located at a distance of 4.16 $\pm$ 0.33~kpc (\citealt{Alvarez2013}, \citealt{Zeidler2015} and also see discussion in \citealt{Tiwari2021}), close to the tangent of the Carina arm at $l$ = 284$\degree$.3, $b$ = - 0$\degree$.3. RCW~49 hosts a compact stellar cluster, Westerlund 2 (Wd2), consisting of 37 OB stars and 30 early-type OB star candidates surrounding it (\citealt{Ascenso2007,Tsujimoto2007,Rauw2011,Mohr-Smith2015,Zeidler2015}). There are two Wolf-Rayet stars also associated with RCW~49: WR20a, which is a part of the Wd2 cluster and is suggested to be among the most massive binaries in the Galaxy \citep{Rauw2005}, and WR20b, which is a few arcminutes away south-east of Wd2. Winds and radiation from these stars are responsible for sculpting the gas and dust into dense molecular clouds, shells and pillars in RCW~49.  

RCW~49 has been studied previously at radio, sub-millimeter, IR and X-ray wavelengths. \citet{Whiteoak1997} and \citet{Benaglia2013} studied radio data (at 0.843, 1.38, 2.38, 5.5, 9~GHz) and reported two shells towards RCW~49. \citet{Furukawa2009} obtained $^{12}$CO (2-1) data with NANTEN2 telescope towards RCW~49. With an angular resolution of 1.5$\arcmin$, they identified two large-scale molecular clouds in velocity ranges of -11 to 9~km~s$^{-1}$ and 11 to 21~km~s$^{-1}$, and suggested that their collision led to the formation of the Wd2 cluster. \citet{Ohama2010} analysed $^{12}$CO (1-0), $^{12}$CO (2-1), and $^{13}$CO (2-1) to estimate temperature and density distributions for the two clouds identified by \citet{Furukawa2009}. \citet{Churchwell2004} studied RCW~49 at mid-IR wavelengths (Infrared Array Camera data at 3.6, 4.5, 5.8, and 8.0~$\mu$m) and identified different regions based on dust emission with respect to the angular radius from Wd2. Moreover, diffuse X-ray observations (0.5--7~keV) probed the hot plasma distributed around Wd2 \citep{Townsley2019}. 

Although the studies mentioned above unveiled the rich region of RCW~49, the observations used for the analysis lacked either spatial or spectral resolution, which hindered precise investigation of the morphology, energetics and physical conditions. The new high resolution \cii\,~158~$\mu$m and the \oi\,~63~$\mu$m observations towards RCW~49 were taken as part of the Stratospheric Observatory For Infrared Astronomy (SOFIA, \citealt{Young2012}) legacy program FEEDBACK\footnote{https://feedback.astro.umd.edu/} \citep{Schneider2020} and the $^{12}$CO observations were taken with the Atacama Pathfinder Experiment (APEX\footnote{APEX is a collaboration between the Max-Planck-Institut f\"{u}r Radio-astronomie,  Onsala  Space  Observatory,  and  the  European  Southern Observatory.}, \citealt{Gusten2006}). In our previous work, we characterised the expanding shell of RCW~49 using the \cii\,, $^{12}$CO and $^{13}$CO observations \citep{Tiwari2021}.    
With the goal of quantifying radiative and mechanical input by massive stars into their surroundings, similar studies were performed to understand the stellar feedback mechanisms in the ISM through the FEEDBACK program (first scientific results in \citealt{Luisi2021,Tiwari2021,Kabanovic2021,Beuther2022}). \\

In this paper, we quantify the effects of stellar feedback on different regions of RCW~49 by determining their physical conditions. We want to understand the contribution of stellar feedback in the evolution of these regions in terms of morphology and future star formation. We compare our observations (obtained through the FEEDBACK program and through the Spitzer telescope) with PDR models to derive the FUV incident flux ($G_{\rm 0}$), density (n), surface temperature ($T_{\rm surf}$) and pressures ($p$) in different regions of RCW~49. Moreover, through this paper, we aim at providing a suitable PDR analysis strategy that can be used to determine the physical conditions in the ISM.

\section{Observations} \label{sec:obs}

\subsection{SOFIA}\label{sec:sofia-obs}

The \cii\ and \oi\ observations were observed during three flights in June of 2019 using upGREAT\footnote{German Receiver for Astronomy at Terahertz. (up)GREAT is a development by the MPI f\"ur Radioastronomie and the KOSMA/Universit\"at zu K\"oln, in cooperation with the DLR Institut f\"ur Optische Sensorsysteme.} \citep{Risacher2018}. The upGREAT receiver can observe both \cii\ and \oi\ lines simultaneously by using a dual 7 pixel low-frequency array (LFA) that was tuned to the \cii\ line, and in parallel a seven pixel high frequency array (HFA) that was tuned to the \oi\ 63 $\mu$m line. 
The observing mode was driven by the \cii\ line sensitivity, thus, limiting the S/N of the \oi\, line. Moreover, the \oi\ data was under sampled (for more details see \citealt{Schneider2020}).

A Fast Fourier Transform Spectrometer (FFTS)
with 4~GHz instantaneous bandwidth and a frequency resolution of 0.244~MHz
\citep{Klein2012} served as backend. Thus, both the \cii\ and \oi\ data have an original resolution of 0.04~km~s$^{-1}$. The data set was then re-binned to a spectral resolution of 0.2~km~s$^{-1}$.
The instrument and telescope optics determine the intrinsic half-power beam widths: 14.1$\arcsec$ for the \cii\ line and 6.3$\arcsec$ for the \oi\ line. All spectra are presented on a main beam brightness temperature ($T_{\rm mb}$) scale. The main beam efficiency ($\eta_{\rm mb}$) values for \cii\ and \oi\ are 0.65 and 0.69, respectively. Further observational details are given \citet{Tiwari2021}.

\begin{figure*}[t]
\centering

\includegraphics[width=180mm]{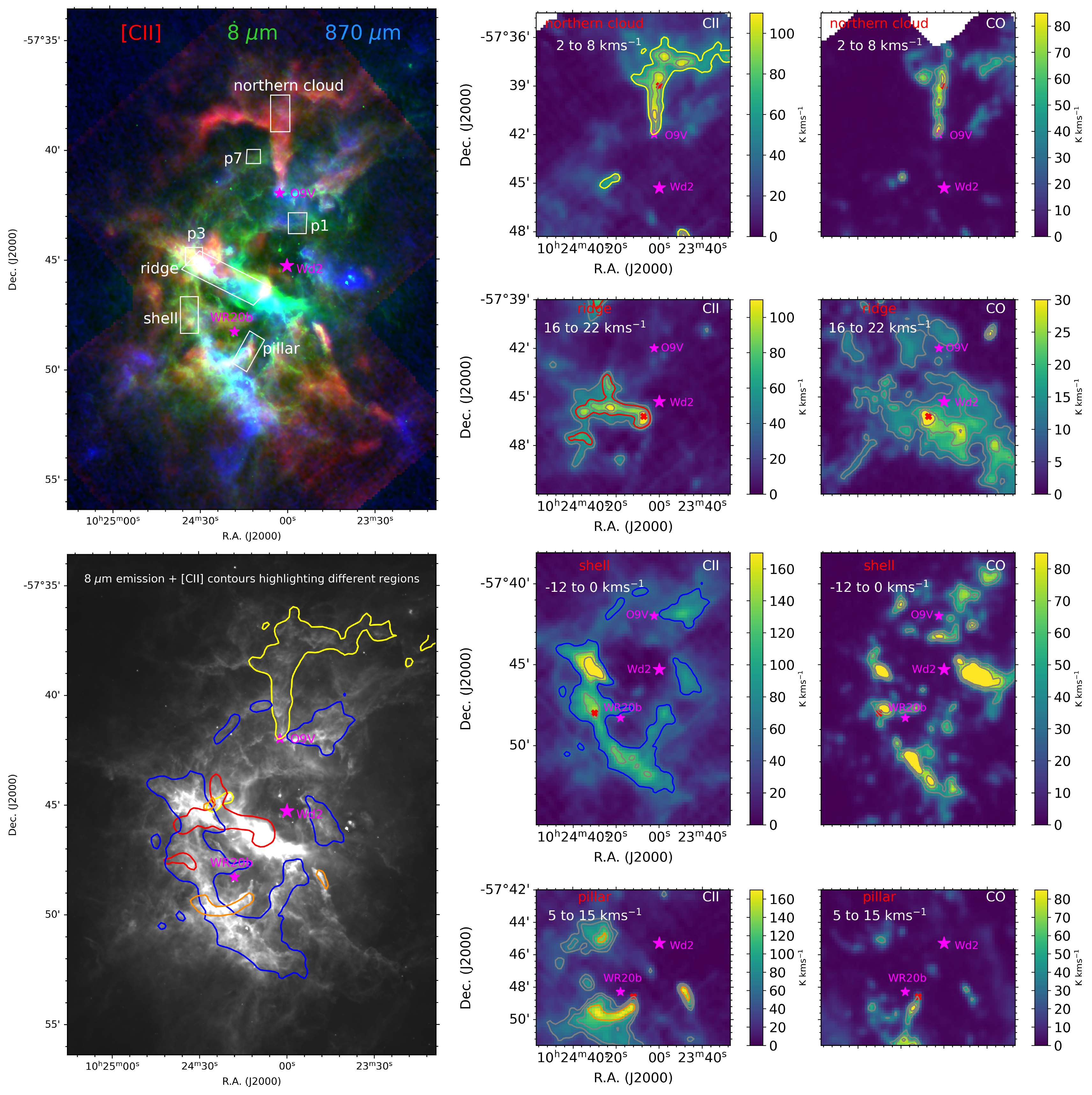}

\caption{A red blue green (RBG) image of RCW~49 is shown in the upper left. Red corresponds to the \cii\ integrated intensity emission within the velocity range of -30 to 30~km~s$^{-1}$. Green corresponds to the GLIMPSE \citep{Benjamin2003} 8~$\mu$m emission and blue corresponds to the ATLASGAL \citep{Schuller2009} 870~$\mu$m continuum. Different interesting regions towards which the PDR analysis is done are shown and outlined in white boxes. The \cii\ and $^{12}$CO emission towards the northern cloud, ridge, shell and pillar are shown in the middle and right panels. The red crosses mark the lines of sight for which the spectra are presented in Fig.~\ref{fig:spectra}. 
In the lower left panel, the 8~$\mu$m emission image is overlaid with \cii\ emission contours highlighting these regions. The contour colors are same to the ones in the \cii\ maps shown in the middle panels.  
The Wd2 cluster, the WR20b star and the O9V star are shown marked in pink in all panels. 
\label{fig:rgb}}
\end{figure*}

\subsection{APEX}\label{sec:apex-obs}
The $J$ = 3-2 transition of $^{12}$CO and $^{13}$CO was observed towards RCW~49 using the LAsMA array on the APEX telescope \citep{Gusten2006}. The beam width is 18.2$\arcsec$ at 345.8~GHz. Advanced FFTS (Klein et al. 2012) are used as backends with a bandwidth of 2 $\times$ 4~GHz and a native spectral resolution of 61~kHz. 
All spectra are calibrated in $T_{\rm mb}$ with $\eta_{\rm mb}$ = 0.68 at 345.8~GHz. After a linear baseline subtraction, all data were binned to 0.2~km~s$^{-1}$. The angular resolution of the maps is $\sim$ 20$\arcsec$. All other data sets were convolved to this resolution for the PDR analysis described later in this paper. For more observational details, see \citep{Tiwari2021}.

\subsection{Spitzer}\label{sec:spitzer-obs}
The H$_2$ observations were carried out using the Infrared Spectrograph (IRS, \citealt{Houck2004}) on board the Spitzer Space Telescope \citep{Werner2004}. 
The H$_2$ 0-0 S(1) and H$_2$ 0-0 S(2) line data presented in this paper are part of the Spitzer program PID 20012 (PI M. Wolfire). The program observed seven fields \citep{Castellanos2014} and we analysed three (p1, p3 and p7) of them as shown in Fig.~\ref{fig:rgb}. The AOR keys of each of them are 13812992, 13813504, and 13814528. The spectrograph observes in four modes depending on the wavelengths and resolution \citep{Houck2004}: short wavelengths with low resolution (SL), long wavelengths with low resolution
(LL), short wavelengths with high resolution (SH), and long wavelengths with high resolution (LH). The S(1) and S(2) lines are at 17.03 and 12.28~$\mu$m observed with the SH mode. The pixel size is 2.26$\arcsec$, the slit size is 4.7$\arcsec$ $\times$ 11.3$\arcsec$, and the mapped regions are $\sim$ $50\arcsec \times 60\arcsec$ for p1, $\sim$ $50\arcsec \times 50\arcsec$ for p3 and $\sim$ $40\arcsec \times 40\arcsec$ for p7.

Observation data cubes (along with uncertainty cubes) were generated with the CUbe Builder for IRS Spectra Maps (CUBISM\footnote{https://irsa.ipac.caltech.edu/data/SPITZER/docs/\\dataanalysistools/tools/cubism/}, \citealt{Smith2007-cubism}), which is a tool used for constructing data cubes of the mapping mode spectra taken with the Spitzer IRS spectrograph. We then used PAHFIT\footnote{http://tir.astro.utoledo.edu/jdsmith/research/pahfit.php} \citep{Smith2007-pahfit} on the data cubes, which decomposes the IRS spectra in order to give us intensity values at each pixel such that we have H$_2$ intensity and uncertainty maps.

\section{Source selection in RCW~49} \label{sec:intro-reg}
Figure~\ref{fig:rgb} shows RCW~49 marked with seven different regions (in white boxes), which are spatially and in some cases are spectrally distinct. 
These regions are at different distances to the Wd2 cluster and some have very distinct morphology, making them ideal to study the impact of stellar feedback in RCW~49.
In this section we introduced these
different regions by describing their physical and chemical
characteristics.

We have selected the ``northern cloud'' as representative of the cloud whose collision with another cloud may have triggered the formation of the Wd2 cluster \citep{Furukawa2009}. This cloud can be spatially disentangled through its velocity range (2 to 8~km~s$^{-1}$).
The ``ridge'' position was selected as part of the other cloud that is thought to have been involved in the cloud collision that gave rise to the formation of Wd2 \citep{Furukawa2009}. Part of the ridge is associated with the inner dust ring identified in the Spitzer study of RCW~49 \citep{Churchwell2004}. The velocity of the ridge is in the range 16 to 22~km~s$^{-1}$.
We have selected a region in the ``shell'' that was identified by \citet{Tiwari2021} as expanding toward us at 12~km~s$^{-1}$. This particular position is well separated from the ridge and pillar positions included in our analysis. The shell spans the velocity range -25 to 0~km~s$^{-1}$ and is most intense  in -12 to 0~km~s$^{-1}$ range.
We have selected the ``pillar'' as it may mark the radiative interaction of Wd2 with the ridge. However, while the pillar points back toward Wd2, we note that the Wolf-Rayet star, WR20b is in close proximity and may have played a key role in its formation and determining its physical conditions.
While the above regions where selected based on morphological considerations of RCW~49, we have added three positions selected from the Spitzer IRS spectral survey also covered by the \cii\,, \oi\ and $^{12}$CO (3-2) surveys. Position ``p1'' (Fig.~\ref{fig:rgb}, upper left panel), marks the transition from the cavity surrounding Wd2 to the bright \cii\ and 8~$\mu$m emission. Position ``p3'' overlaps with the eastern part of the ridge. Position ``p7'' is at the northern tip of the expanding shell identified by \citet{Tiwari2021}.

\section{Spatial distribution of observed species}\label{sec:spatial-dist}
The northern cloud has bright \cii\ and $^{12}$CO emission but appears to be less prominent in GLIMPSE 8~$\mu$m emission (Fig.~\ref{fig:rgb}). This is likely due to the high spectral resolution of the \cii\ and $^{12}$CO (3–2) observations, which allows us to disentangle the emission from this component (from 2 to 8~km~s$^{-1}$) from the surroundings. 
Figure~\ref{fig:rgb} shows bright \cii\,, $^{12}$CO and 8~$\mu$m emission towards the ridge. 
The shell is outlined very well by the \cii\ and $^{12}$CO emission as seen in Fig.~\ref{fig:rgb}. However, unlike the \cii\ emission distribution, the $^{12}$CO emission distribution is fragmented towards the shell. This is also seen in the $^{13}$CO emission map towards this region \citep{Tiwari2021}. For the 8~$\mu$m emission, similar to the northern cloud, one cannot disentangle the emission solely from the shell due to its lack of velocity resolution. 
Bright \cii\,, $^{12}$CO and 8~$\mu$m emission is observed towards the pillar. 

\begin{figure}[h]
\centering
\includegraphics[width=0.48\textwidth]{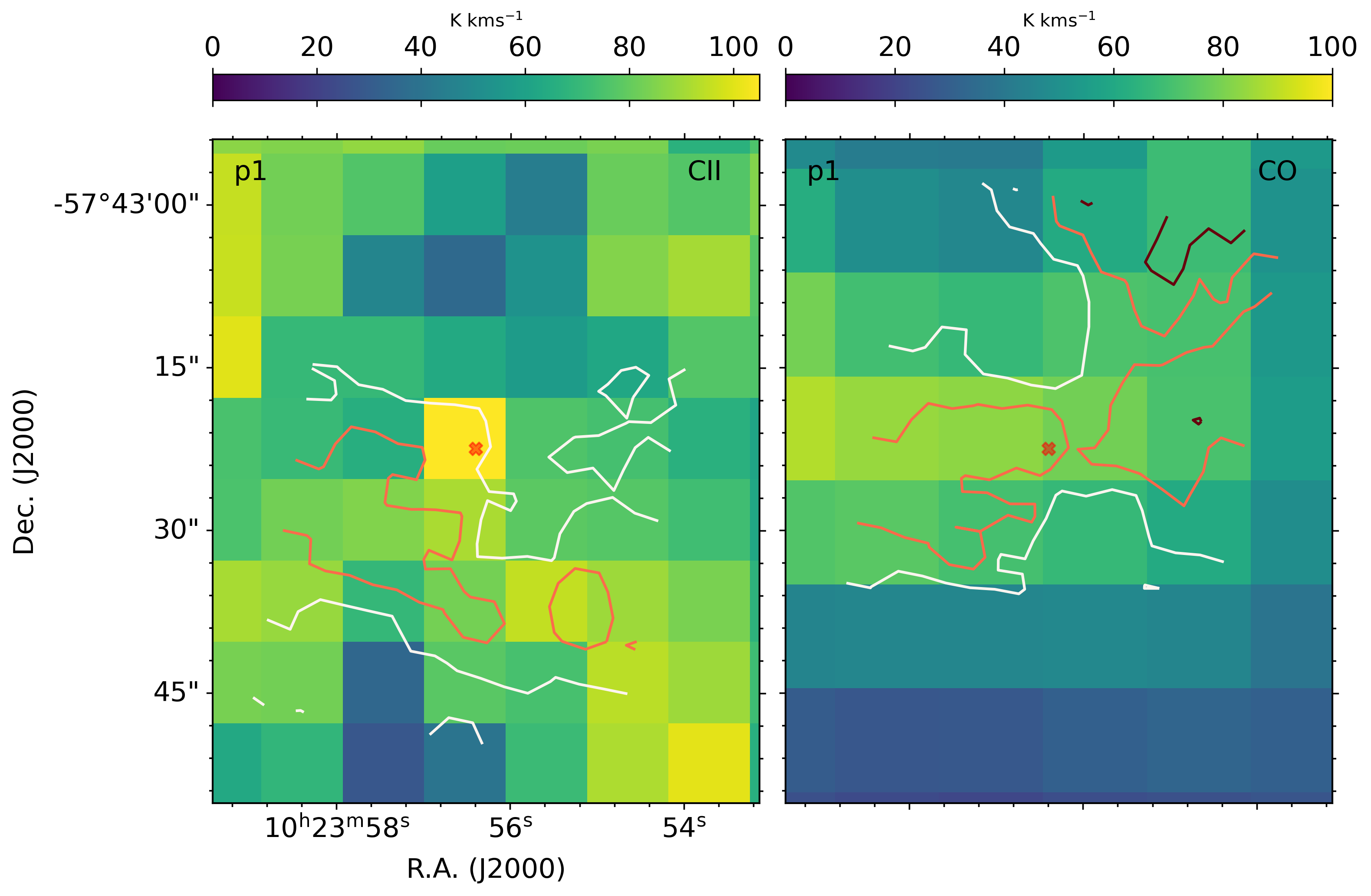}
\includegraphics[width=0.48\textwidth]{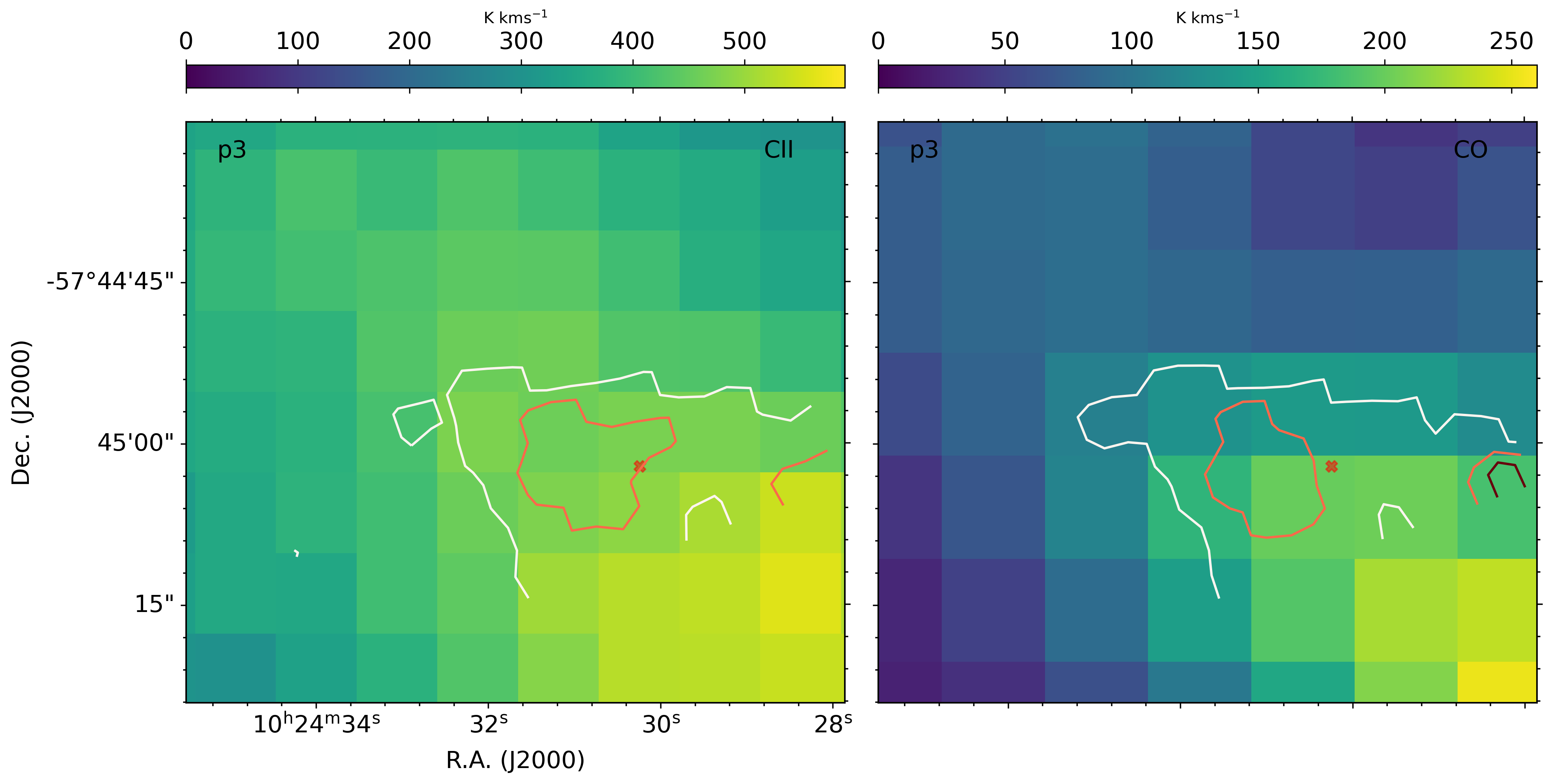}
\includegraphics[width=0.48\textwidth]{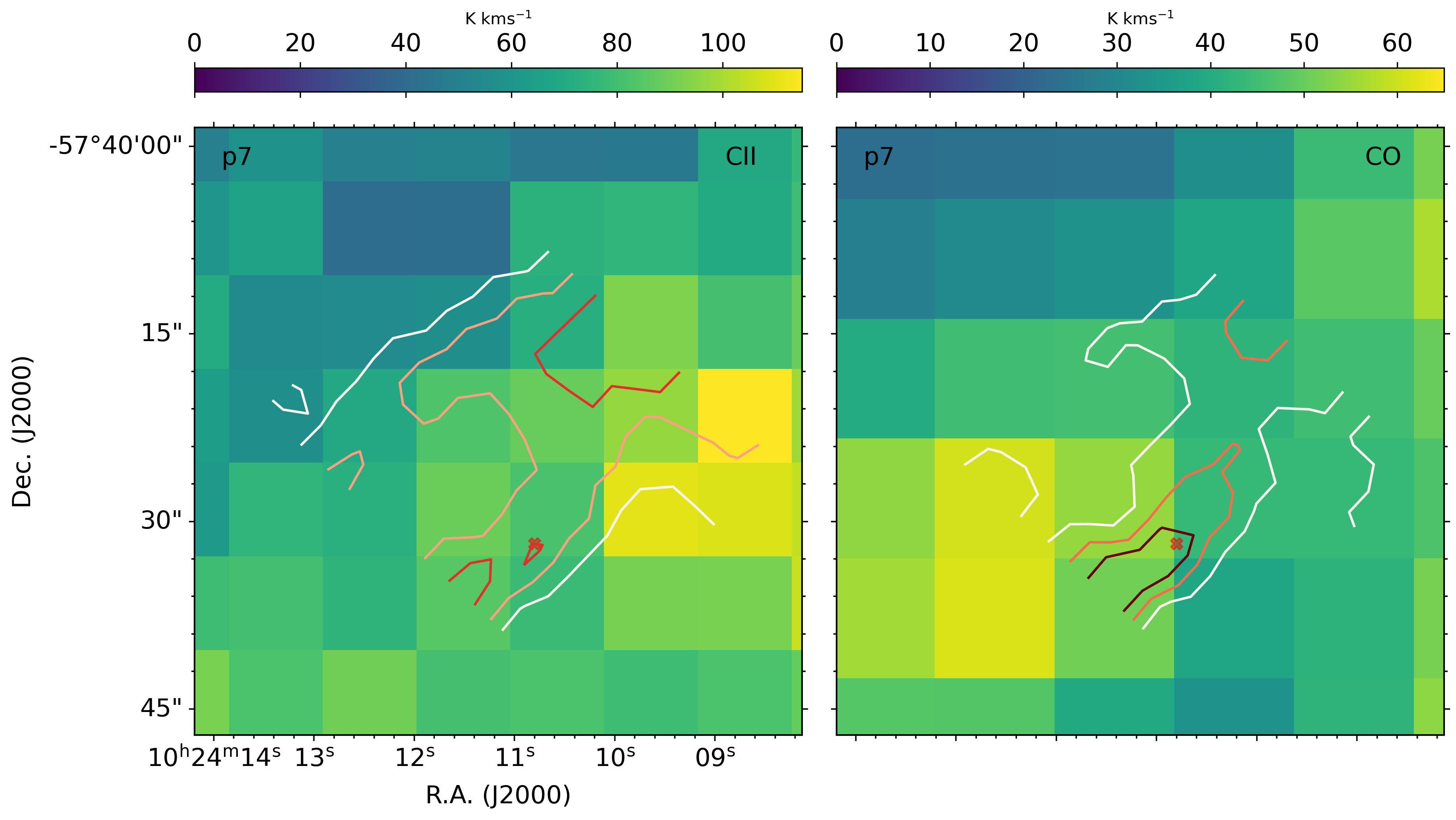}
\caption{Emission maps of \cii\ and $^{12}$CO (3--2) towards the regions p1, p3 and p7 (shown in Fig.~\ref{fig:rgb}). 
The \cii\ maps are overlaid with H$_2$ 0-0 S(1) intensity contours and the $^{12}$CO maps are overlaid with  H$_2$ 0-0 S(2) intensity contours. The contours are colored such that the darkest shade of red corresponds to the highest intensity. The red crosses mark the lines of sight for which the spectra are presented in Fig.~\ref{fig:spectra}. \label{fig:h2maps}}
\end{figure}

The \cii\ and $^{12}$CO (3-2) emission maps overlaid with H$_2$ 0-0 S(1) and H$_2$ 0-0 S(2) emission contours respectively, towards these positions are shown in Fig.~\ref{fig:h2maps}. Due to the lower velocity resolution 
$\Delta v \sim 500$ km ${\rm s^{-1}}$
in the IR line data, we compare their spatial distribution with the \cii\ and $^{12}$CO maps where their emission is integrated over the entire velocity range. This ensures a fair comparison between the IR, FIR and submm wavelength data presented here.
In general, the H$_2$ 0-0 S(1) emission distribution seems to be related more to the \cii\ emission distribution than to the $^{12}$CO emission distribution, and the H$_2$ 0-0 S(2) emission distribution seems to be related more to the $^{12}$CO emission distribution than to the \cii\ emission distribution. This suggests that the H$_2$ 0-0 S(1) line traces less dense gas compared to the H$_2$ 0-0 S(2) line.
For p1, there are differences in the spatial distribution of the emission between H$_2$ 0-0 S(1) and H$_2$ 0-0 S(2). 
In p3, the H$_2$, \cii\ and $^{12}$CO emission is brighter in the south-west, which is in the direction of the Wd2 cluster. 
In p7, the H$_2$ emission is essentially tracing the northern-most part of the shell. This structure is less detailed in the \cii\ and $^{12}$CO maps because of the lower angular resolution compared to the IR data.  
The dependence of the H$_2$ lines' emission on the physical conditions can be seen in PDR model contour plots\footnote{e.g., https://dustem.astro.umd.edu/models/wk2006/h200s2s1\\z1web.html}. Along with the H$_2$ lines, several ionised gas tracers ([\siv] 10.5\,$\mu$m, [\neii] 12.8\,$\mu$m, [\neiii] 15.6\,$\mu$m, [\siii] 18.7\,$\mu$m) were also mapped using the Spitzer telescope. Their corresponding IR spectra averaged over the regions of p1, p3 and p7 are discussed and shown in Appendix~\ref{app:ir-spec} and Fig.~\ref{fig:ir-spec}, respectively.

\begin{figure}
\centering
\includegraphics[width=75mm]{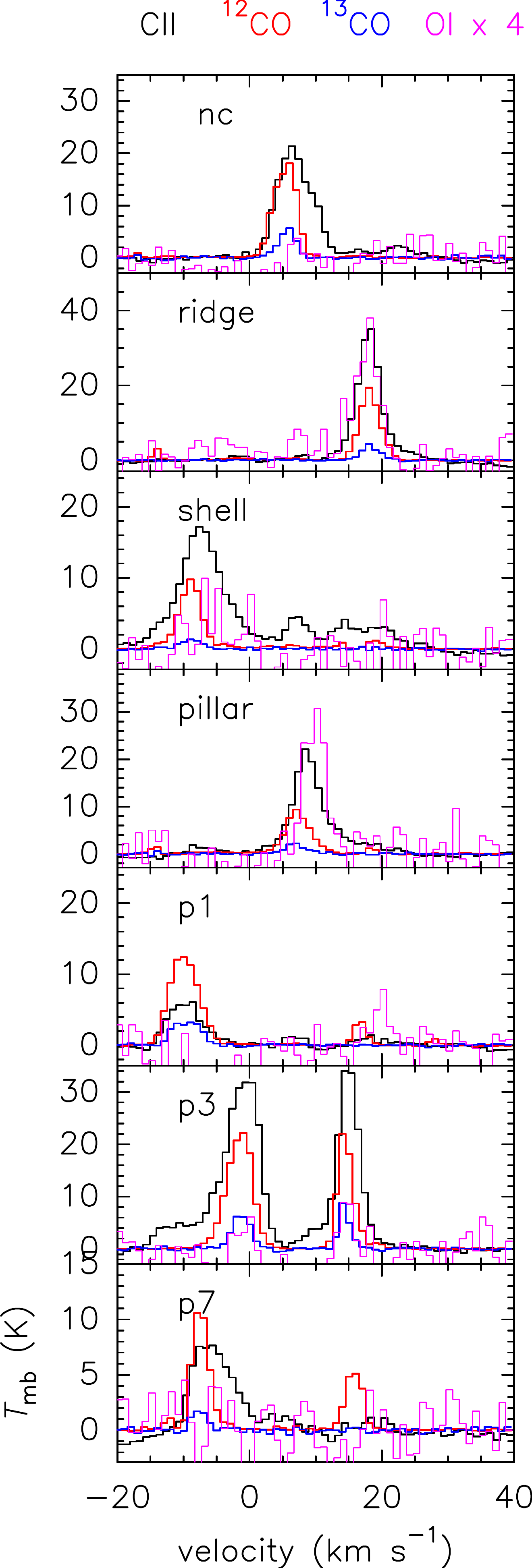}

\caption{Representative spectra of \cii\,, $^{12}$CO, $^{13}$CO and \oi\ towards the northern cloud (nc), ridge, shell, pillar, p1, p3 and p7. All data was convolved to the same angular resolution of $\sim$ 20$\arcsec$ and the spectral resolution of the spectra shown here is 1~km~s$^{-1}$. 
\label{fig:spectra}}
\end{figure}

\section{Spectra}
\label{sec:spectra}
Figure~\ref{fig:spectra} shows the representative spectra of the \cii\,, $^{12}$CO, $^{13}$CO and \oi\,~63~$\mu$m lines towards the seven regions analysed in detail in this work.

We have convolved the maps to the same spatial resolution and extracted data for a single pixel. The similarity in peak position and profile of all tracers for positions, the ridge, p1 and the two velocity components in p3, ensures that we are tracing the same gas.  However for the northern cloud, shell, pillar and p7, the
velocity peak positions of the \cii\ and \oi\ lines are shifted by $\sim$ 1~km~s$^{-1}$ with respect to the $^{12}$CO line.
These velocity shifts are in accordance with expectations for the advection of gas through the PDR \citep{Tielens2005} and we surmise that these species trace gas in the same PDR. In our source selection, we avoided clumps, which is vital for the PDR analysis described in Sect.~\ref{sec:overlay}.
Though not used in our PDR analaysis, we also present spectra of $^{13}$CO towards the different regions. We can see that the  $^{13}$CO spectral profile follows that of $^{12}$CO, suggesting no self absorption in $^{12}$CO from any cold foreground. Similarly, we did not observe any self absorption features in the \oi\ line profiles along the lines of sight where it was detected. This is in contrast with most \oi\ observations towards other Galactic sources (\citealt{Poglitsch1996,Liseau2006,Gerin2015,Rosenberg2015,Schneider2018,Goldsmith2021}).

We considered the intensities integrated over the specific velocity ranges corresponding to their regions (same as shown in Fig.~\ref{fig:rgb}). For the northern cloud and p1, the \oi\ intensity values are upper limits to the \oi\ emission towards these regions. For \cii\,, $^{12}$CO and \oi\,, they are estimated for the entire mapped regions by finding the rms noise in an emission free velocity window. Their values are given in Table~\ref{tab:int-exp}. 
Furthermore, there are also calibration uncertainties, which could be up to 10\% for $^{12}$CO data (APEX, LASMA), $<$ 20\% for the \cii\ data (SOFIA, upGREAT) and $\geq$ 20\% for the \oi\ data.

\begin{figure*}[t]
\centering
\includegraphics[height=0.231\textwidth]{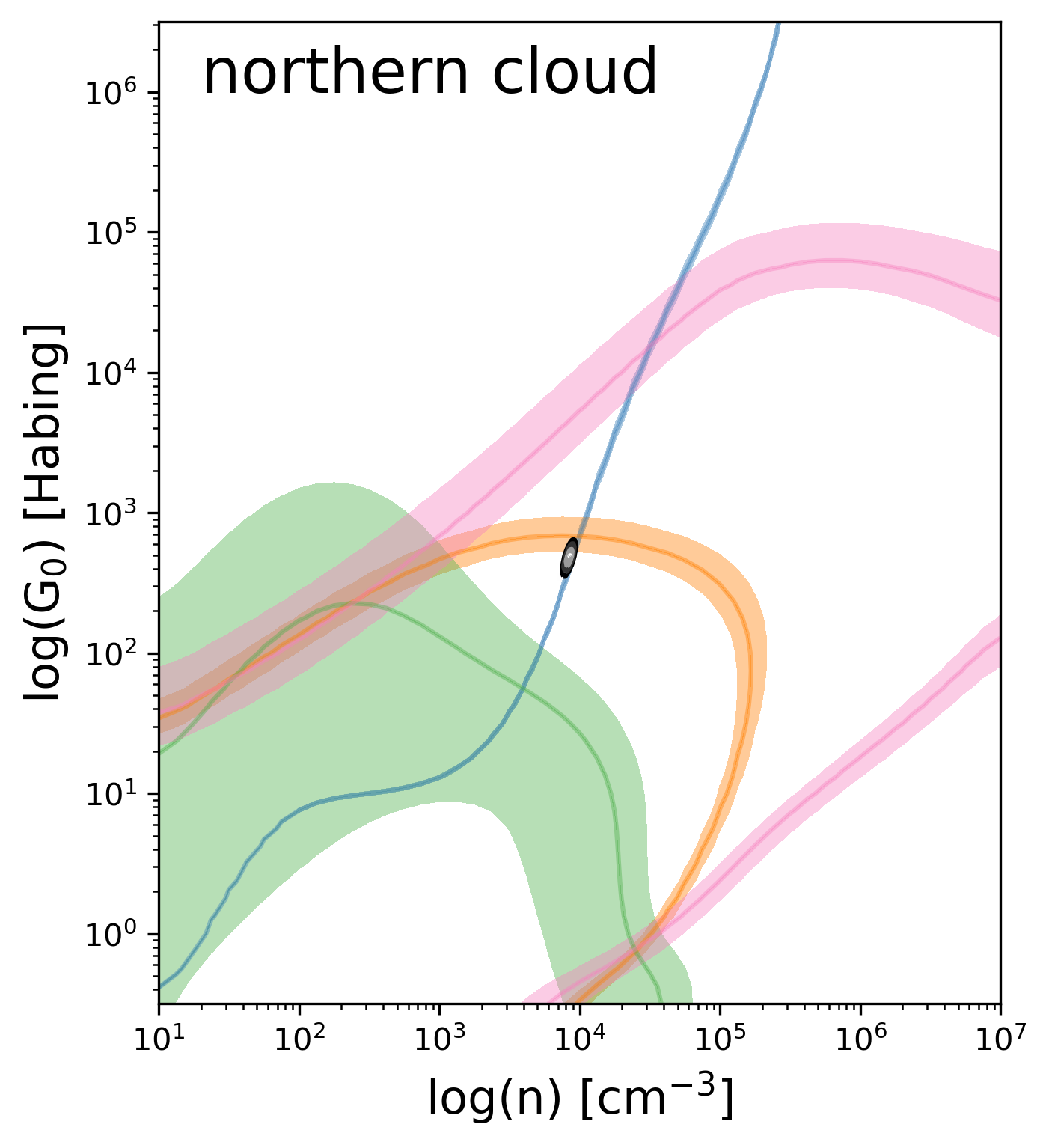}
\includegraphics[height=0.231\textwidth]{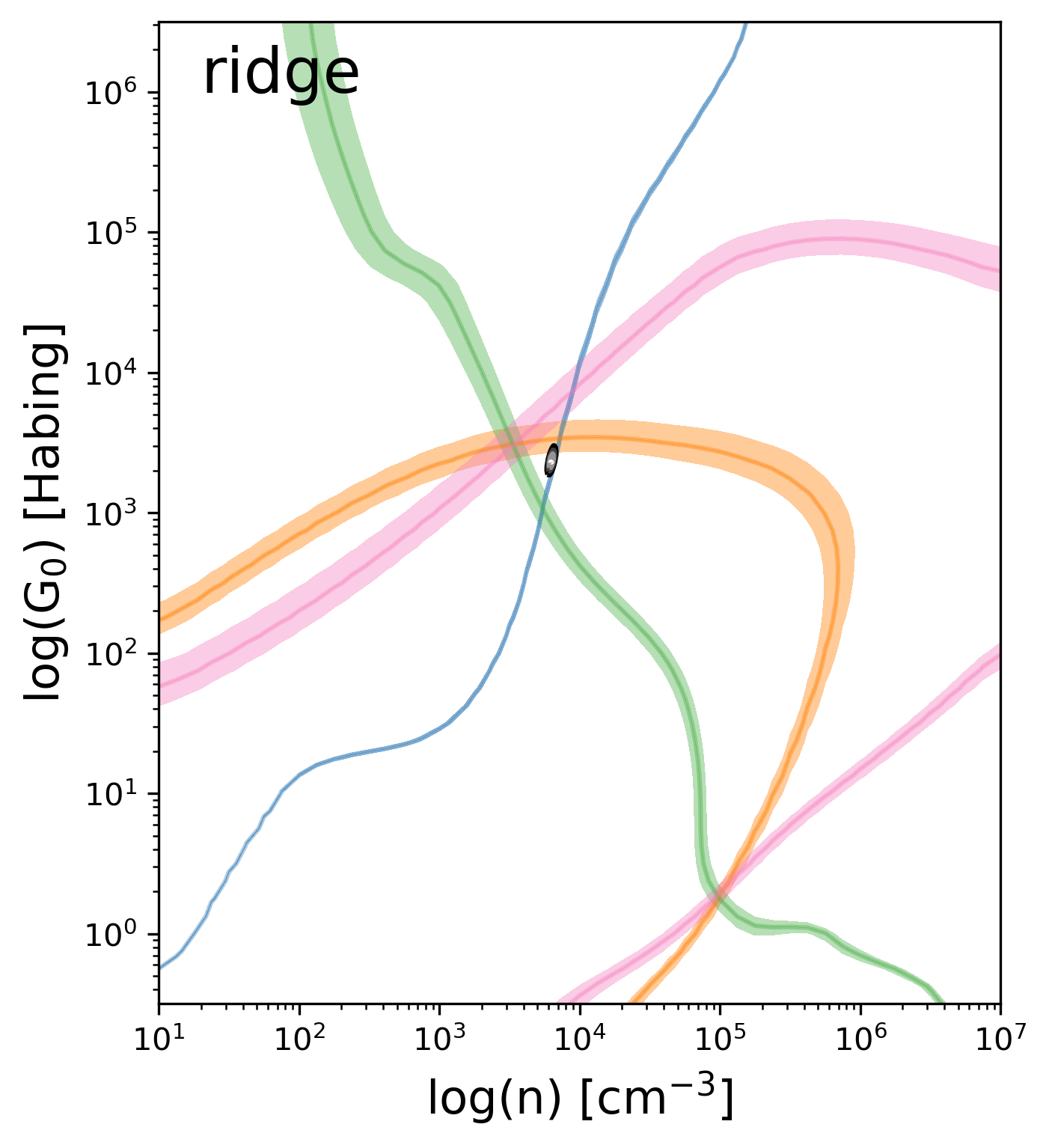}
\includegraphics[height=0.231\textwidth]{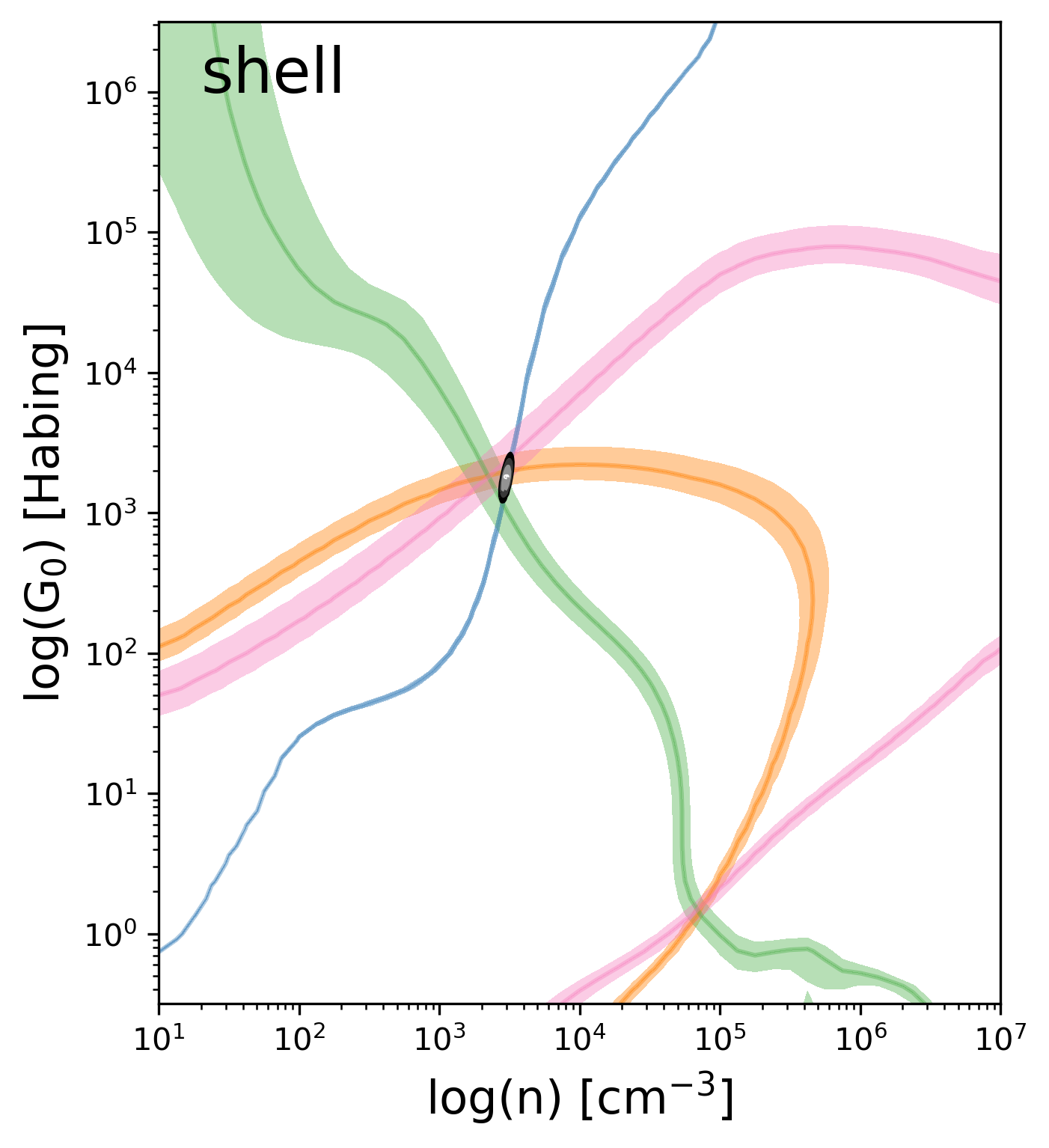}
\includegraphics[height=0.231\textwidth]{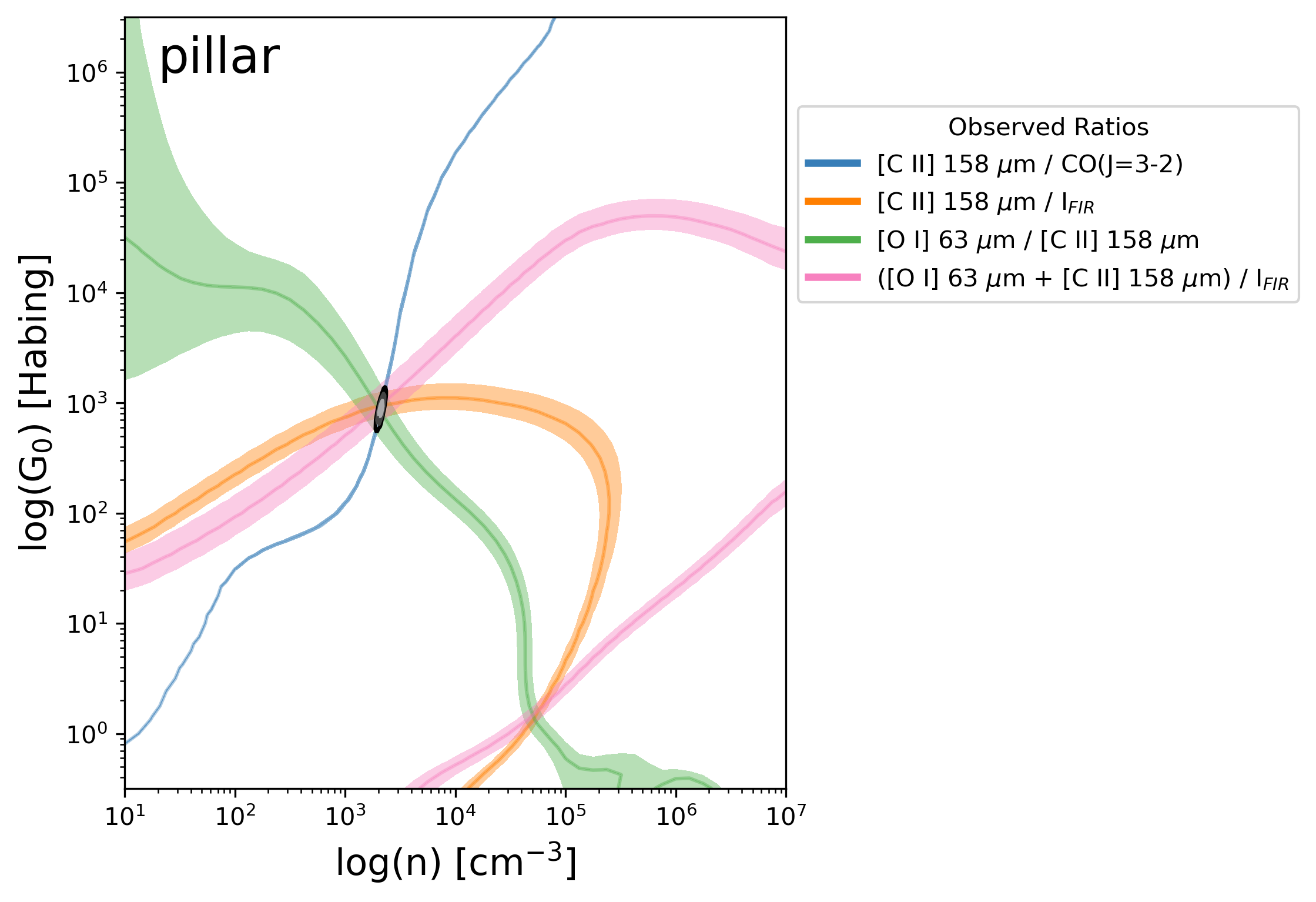}

\caption{Overlay plots of the observed intensity ratios in the PDR Toolbox modeled grid of $G_{\rm 0}$ and $n$ for the northern cloud, ridge, shell and pillar. The shaded region corresponds to the error on the observations. Each plot is overlaid with the contours of its corresponding corner plot
.    \label{fig:spaghetti-4-regs}}

\end{figure*}

\begin{figure*}[t]
\centering
\includegraphics[height=0.24\textwidth]{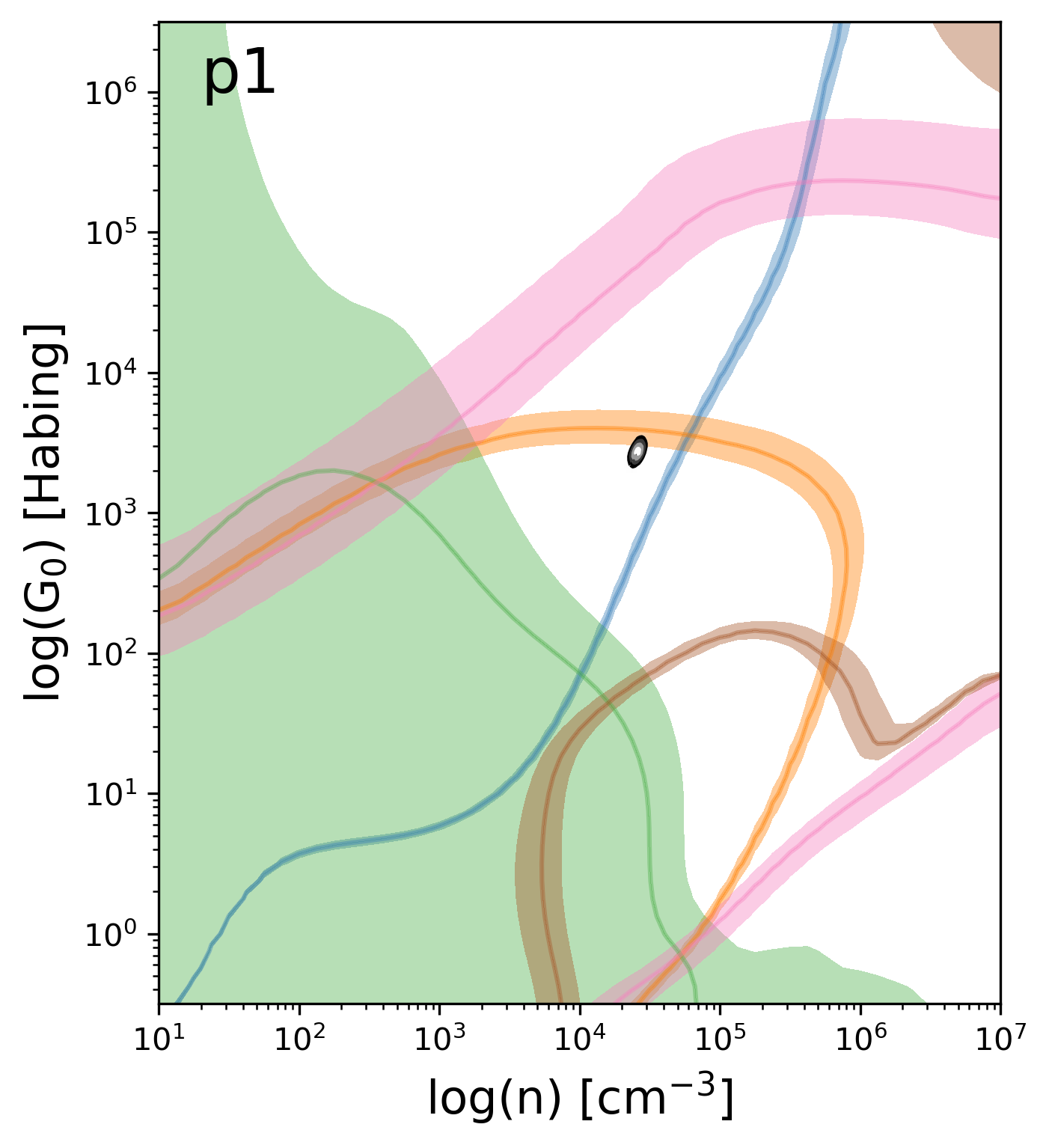}
\includegraphics[height=0.24\textwidth]{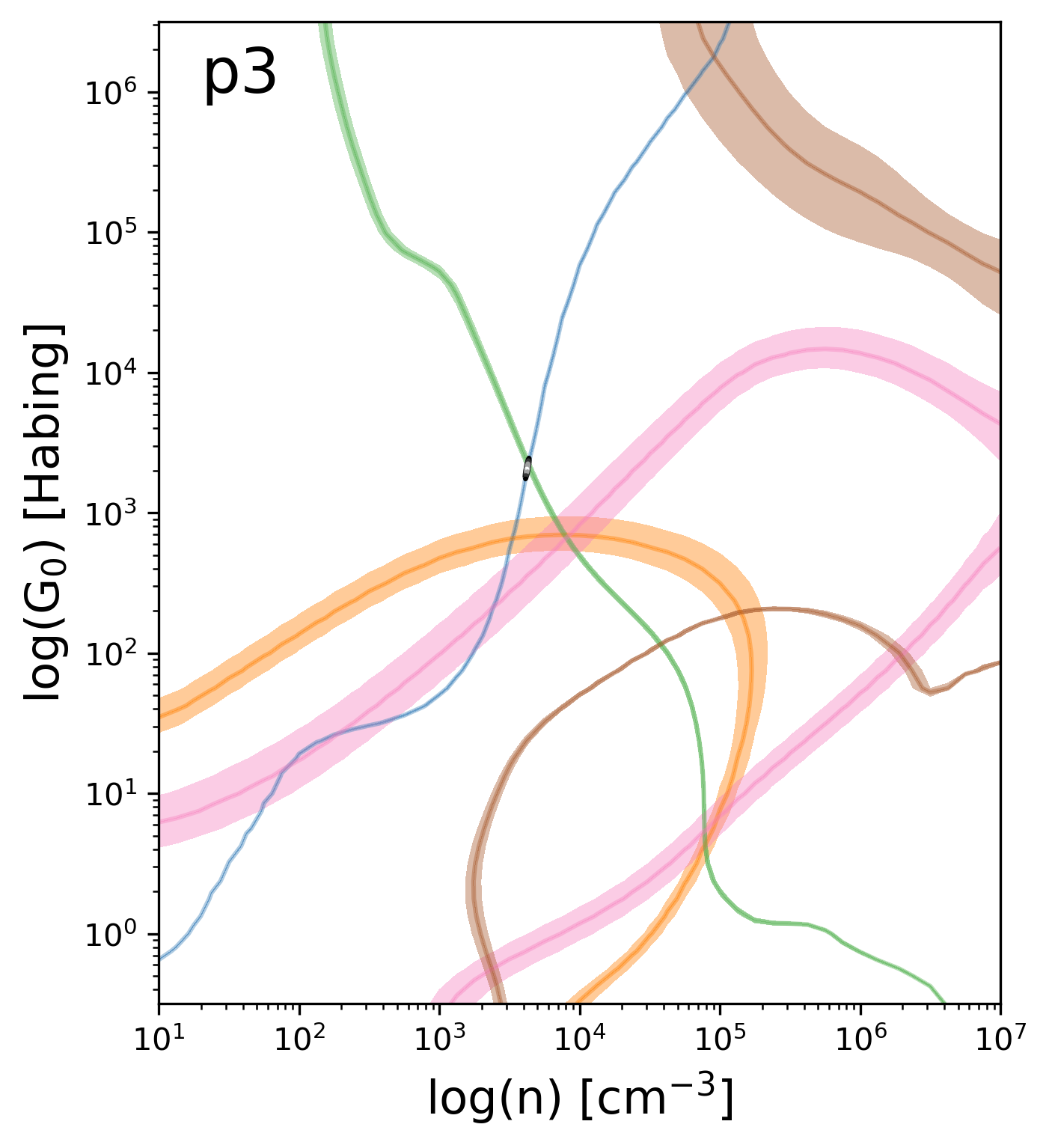}
\includegraphics[height=0.24\textwidth]{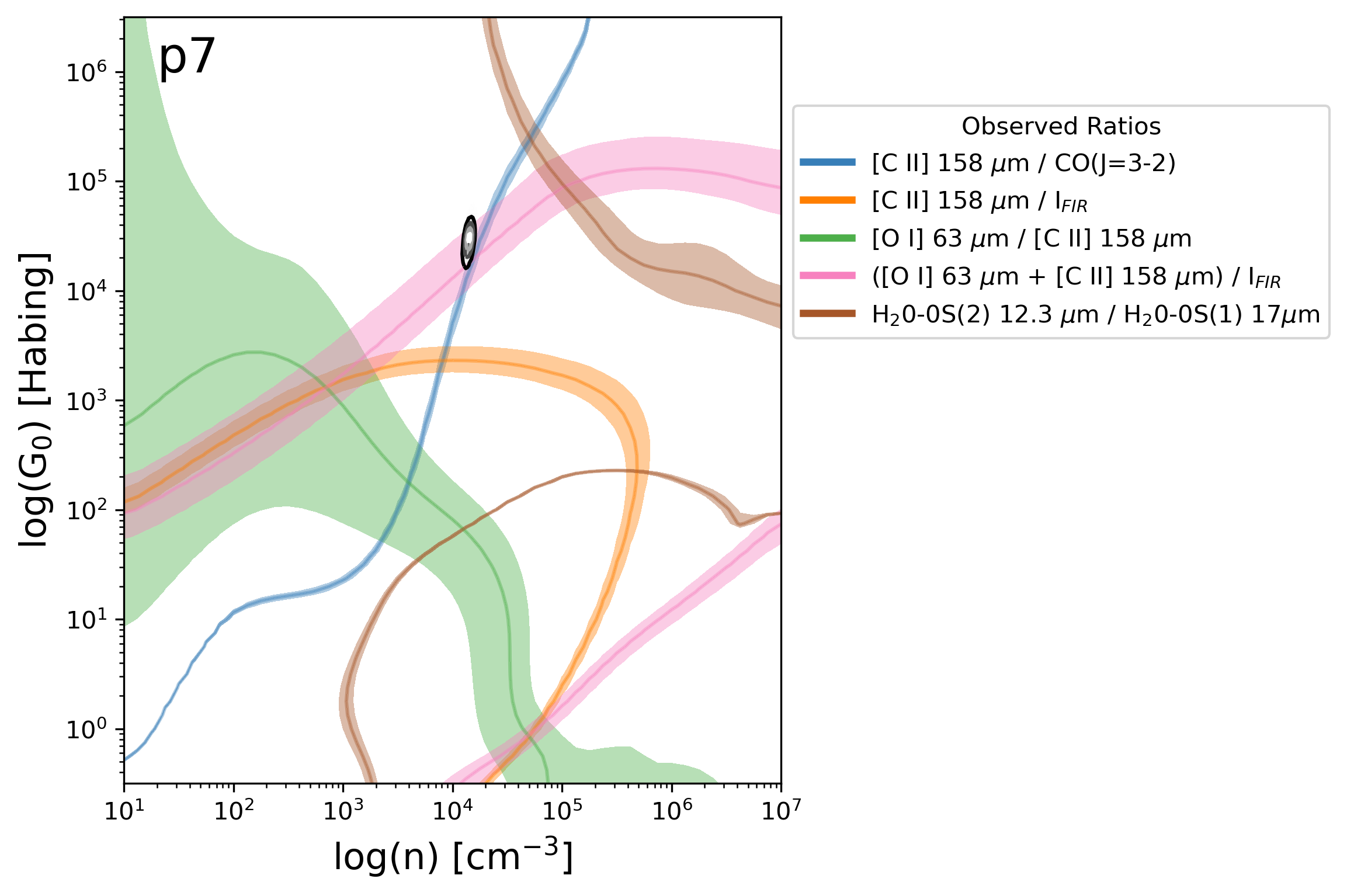}

\includegraphics[height=0.24\textwidth]{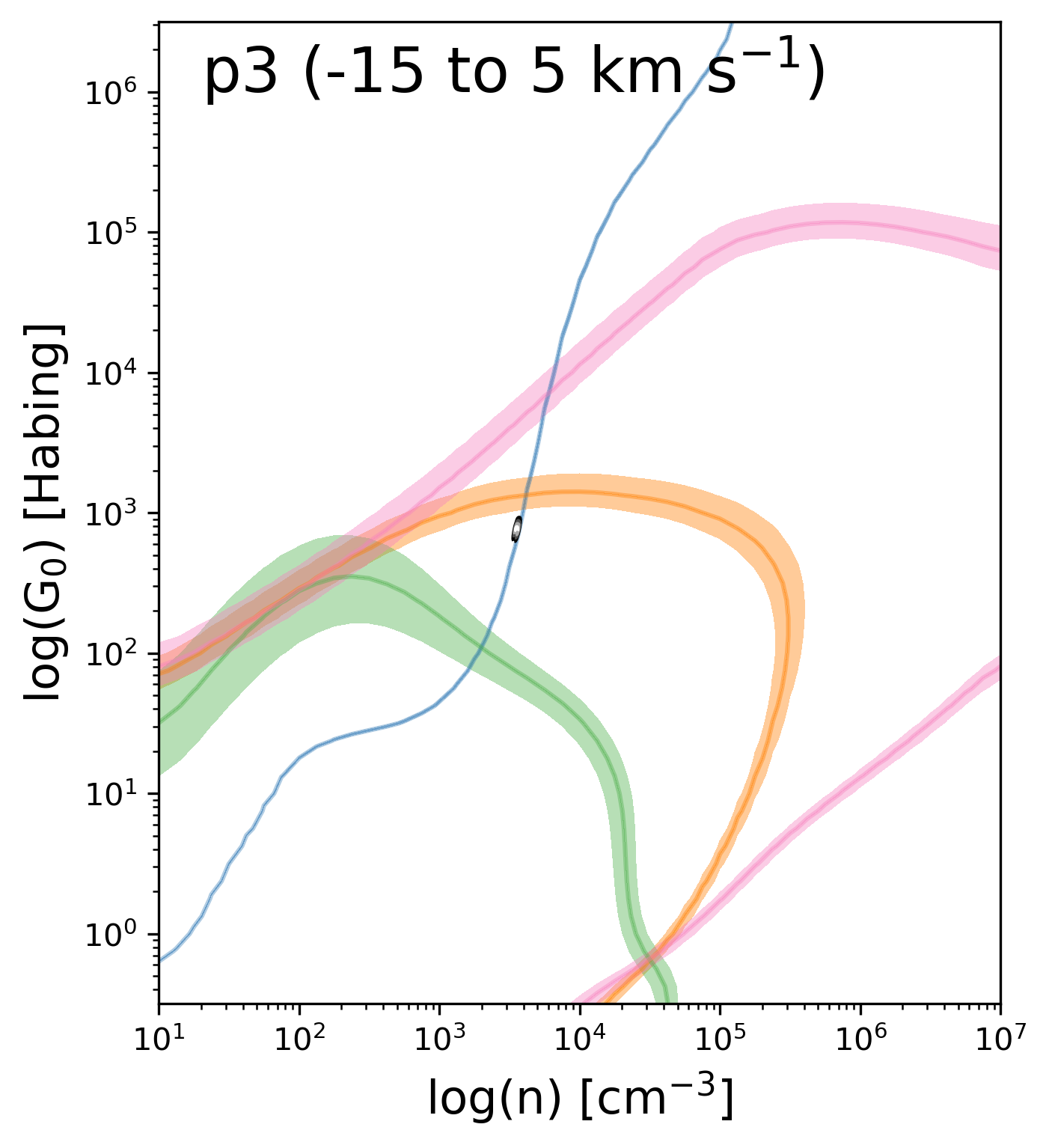}
\includegraphics[height=0.24\textwidth]{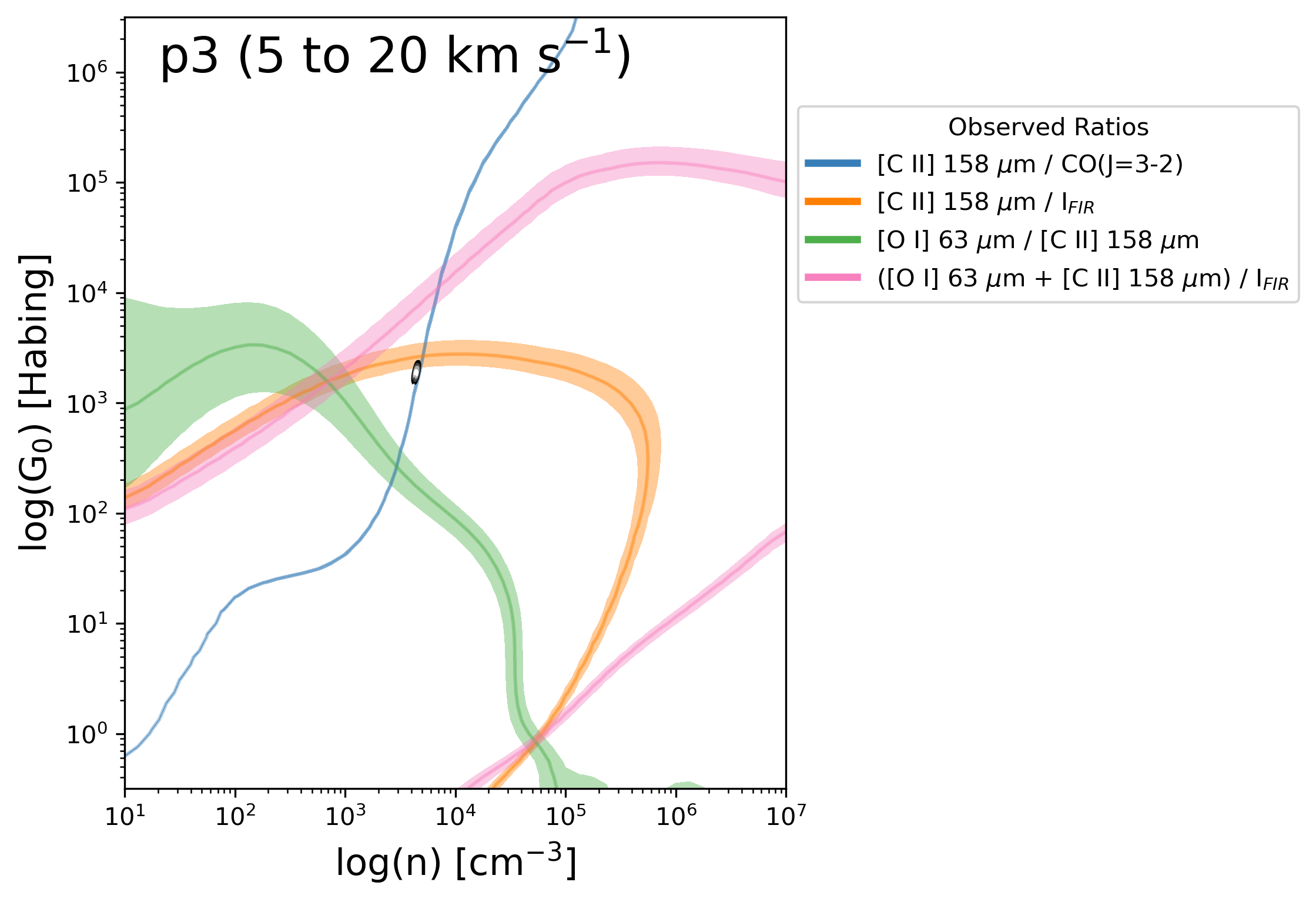}

\caption{Overlay plots of the observed intensity ratios in the PDR Toolbox modeled grid of $G_{\rm 0}$ and $n$ for p1, p3 (upper and lower panels include and exclude H$_2$ data respectively.) and p7 in the upper panel.
The shaded region corresponds to the error on the observations. Each plot is overlaid with the contours of its corresponding corner plot 
.    \label{fig:spaghetti-h2}}
\end{figure*}

\section{PDR diagnostics}
\label{sec:pdr-diag}

We have used the PDR Toolbox\footnote{http://dustem.astro.umd.edu/} (\citealt{Kaufman2006,Pound2008,Pound2011}), a Python-based software package which employs state-of-the-art PDR models to determine the physical conditions in PDRs (the far UV flux ($G_{\rm 0}$) and H nucleus density ($n$)) from observations.

\begin{deluxetable*}{l c c c c c c c c c c c c}
\tablecaption{Observed and model intensity$^*$ values in different regions of RCW~49.}
\label{tab:int-exp}
\tablehead{
 \colhead{Parameter} & 
 \colhead{rms noise} &
 \colhead{northern cloud} & 
 \colhead{ridge} & 
 \colhead{shell} & 
 \colhead{pillar} & 
 \colhead{p1} &
 \colhead{p3} & 
 \colhead{p3v1} &
 \colhead{p3v2} &
 \colhead{p7}
}

\startdata
{R.A. (J2000)} & & 10:23:59.7 &10:24:06.5 &  10:24:29.7 & 10:24:11.7 & 10:23:56.2 & 10:24:30.3 & & & 10:24:11.1\\
{Dec. (J2000)} & & -57:38:52.7 & -57:46:17.2&-57:48:12.4   &-57:49:06  &-57:43:22.7  &-57:45:02.5  & & & -57:40:32.3\\ \hline
 {$I$(\cii\,)} & 0.18 & 6.7  & 9.9   & 9.4   & 11.0   & 3.0   & 34.0 &20.0 &12.0   & 4.2\\
 {$I$($^{12}$CO (3-2))} & 0.003 & 0.032  & 0.029   & 0.017   & 0.017   & 0.028   & 0.076 & 0.047 & 0.029 & 0.014\\
 {$I$(\oi\,)}  &3.3 & 3.7  & 30.0   & 20.0   & 18.0   & 3.3   & 110.0 & 13.0 & 15.0 & 14.0\\
  {$I$(H$_2$ 0-0 S(1))} & 0.05 & & & & & 0.31 & 1.6  & & & 1.0  \\ 
{$I$(H$_2$ 0-0 S(2))} & 0.06 & & & & & 0.75  & 2.2  & & &1.2  \\
{$I$(FIR)}& & 2400 & 12200 & 8200 & 5700 & 4200 & 12400 & & &3800 \\ \hline
{$I_{\rm pre}$(\cii\,)} &  & 4.0 & 6.2 & 5.5 & 4.6 & 6.6 & 5.9 & 4.7& 5.6& 10.0 \\
{$I_{\rm pre}$($^{12}$CO (3-2))} &  & 0.023 & 0.021 & 0.011 & 0.0082 & 0.039 & 0.014 & 0.013& 0.014& 0.031 \\
{$I_{\rm pre}$(\oi\,)} &  & 11.0 & 24 & 14 & 8.7 & 67.4 & 17 & 11& 17& 110 \\
{$I_{\rm pre}$(H$_2$ 0-0 S(1))} &  & & & & & 0.79  & 0.34  & & &0.220  \\
{$I_{\rm pre}$(H$_2$ 0-0 S(2))} & & & & & & 0.16  & 0.022  & & &0.75  \\
\enddata
\tablecomments{* All intensities are in the units of 10$^{-4}$ erg cm$^{-2}$ s$^{-1}$ sr$^{-1}$. The observed intensities were converted from K~km~s$^{-1}$ ($W$) to erg~cm$^{-2}$~s$^{-1}$~sr$^{-1}$ ($I$). 
For \cii\,~158~$\mu$m, $^{12}$CO (3--2) and \oi\,~63~$\mu$m, respectively, the conversions are $I$ (erg~s$^{-1}$~cm$^{-2}$~sr$^{-1}$) = 7.05 $\times$ 10$^{-6}$ $W$ (K~km~s$^{-1}$), 4.238 $\times$ 10$^{-8}$ $W$ (K~km~s$^{-1}$) and 1.1 $\times$ 10$^{-4}$ $W$ (K~km~s$^{-1}$).
$I$ and $I_{\rm pre}$ are the observed intensities and intensities predicted by models, respectively. The columns p3, p3v1 and p3v2 correspond to the intensities integrated over the velocity window of -30 to 30~km~s$^{-1}$, -15 to 5~km~s$^{-1}$ and 5 to 20~km~s$^{-1}$, respectively.
}
\label{tab:int-exp}
\end{deluxetable*}


In fitting the observations, we used the ``Wolfire-Kaufman 2020 (wk2020)'' model set. These are plane-parallel PDR models, with radiation that is incident on one side normal to the layer, that solve for the gas temperature in thermal equilibrium, and atomic and molecular abundances in steady state.
This model set has updated chemistry from that used in 
the \cite{Kaufman2006} models. Most notably, the chemical rates,  PAH chemistry, and 
collision rates  discussed in \cite{Hollenbach2012}, \cite{ Neufeld2016},
\cite{Kovalenko2018}, \cite{Tran2018},
and \cite{Dagdigian2019}, 
plus photodissociation and photoionization rates in \cite{Heays2017}, 
and collisional excitation rates for \oi\ in \cite{Lique2018}. The models are carried out to a depth $A_{\rm v}=7$ with grain-surface chemistry turned off. We also adopt an incident (primary) cosmic-ray ionization rate per hydrogen $=2.0\times 10^{-16}$ ${\rm s^{-1}}$ that decreases with depth as $1/N$ as suggested by \cite{Neufeld2017}. The PDR Toolbox uses the face-on line intensities that emerge from the PDR. 

We used the \cii\,, \oi\ and $^{12}$CO (3-2) observations in the northern cloud, ridge, shell and pillar. In addition to the above species, we used the H$_2$ 0-0 S(1) and H$_2$ 0-0 S(2) lines in p1, p3 and p7. The models take into account optical depth effects in the PDR. Moreover, we do not see any signature of self-absorption in our spectra (Fig.~\ref{fig:spectra}) from a cold foreground. 
We convolved all the observations to the same angular resolution of $^{12}$CO, which is $\sim$ 20$\arcsec$. The observed line intensities are summarized in Table~\ref{tab:int-exp}. As mentioned before, these values are toward specific lines of sight (see Fig.~\ref{fig:rgb} and \ref{fig:h2maps}).

\subsection{Overlay plots}\label{sec:overlay}
Plotting overlays of intensity ratios of the observed species in the model space is a good way to understand the phase space (Figs.~\ref{fig:spaghetti-4-regs} and \ref{fig:spaghetti-h2}) of allowable solutions. In such plots, locations where many observational lines cross indicate areas of ($n,G_0$) space that provide a good fit to the observations. 
For p1 and p7, we used the \cii\,, \oi\,, and $^{12}$CO integrated intensities of the -15 to -5~km~s$^{-1}$ and the -12 to 0~km~s$^{-1}$ components, respectively, as these dominate the spectra (Fig.~\ref{fig:spectra}) and the H$_2$ data. 
However, for p3, both -15 to 5 and 5 to 20~km~s$^{-1}$ velocity components contribute significantly to the emission. Thus, we carried out the PDR analysis for p3 for three cases: first, where we used the integrated intensities of \cii\,, \oi\ and $^{12}$CO in the entire velocity spread of -30 to 30~km~s$^{-1}$ together with the H$_2$ emission; second, where we only used the integrated intensities of \cii\,, \oi\ and $^{12}$CO within the -15 to 5~km~s$^{-1}$ velocity range and third, where again, we only used the integrated intensities of \cii\,, \oi\ and $^{12}$CO within the 5 to 20~km~s$^{-1}$ velocity range.
Moreover, we included ratios of the observed line intensities to the FIR continuum intensity. The FIR intensities were determined using the formalism described in \citet{Goicoechea2015}, where the dust parameters (dust temperature and opacity) were calculated from 70 and 160~$\mu$m data from the Herschel Space Archive (HSA), obtained within the Hi-GAL Galactic plane survey \citep{Molinari2010} observed with the Photodetector Array Camera and Spectrometer (PACS; \citealt{Poglitsch2010}). A detailed description of the method is given in \citep{Tiwari2021}.

\begin{deluxetable*}{l c c c c c c c c c c}
\tablecaption{Derived physical parameters in different regions of RCW~49.}
\label{tab:phys-val-mwp}
\tablehead{
 \colhead{Parameter} & 
 \colhead{units} & 
 \colhead{northern cloud} & 
 \colhead{ridge} & 
 \colhead{shell} & 
 \colhead{pillar} & 
 \colhead{p1} &
 \colhead{p3} & 
 \colhead{p3v1} & 
 \colhead{p3v2} & 
 \colhead{p7}
}
\startdata
{$n$} & 10$^3$~cm$^{-3}$ & 8.6(0.5) & 6.4(0.35)  & 3.1(0.2)  & 2.2(0.1)  & 27.0(2.0) & 4.3(0.1) & 3.6(0.1) &4.5(0.2) &14.0(1.0) \\
{$G_{\rm 0}$} & 10$^3$ Habing units & 0.5(0.07) & 2.4(0.3) & 1.9(0.4) & 0.97(0.21) & 2.9(0.32) & 2.1(0.2) & 0.79(0.07) & 2.0(0.2) &32.0(8.0)\\
{$G_{\rm 0}$(FIR)} & 10$^3$ Habing units& 0.92 & 4.7 & 3.2 & 2.2  & 1.7  & 4.8 & & & 1.5 \\
{$T_{\rm surf}$} & K & 160 & 246 & 250 & 223 & 244 & 242 & 196 & 242 & 441 \\
$T_{\rm ex}$(\cii\,) & K & 55.2 & 72.0 &47.3&54.6&33.6& &70.4&73.0&36.8\\
$p_{\rm th}$ & $10^6$ K~cm$^{-3}$ & 1.4 & 1.6 & 0.77 & 0.5 & 6.6 & 1.0 & 0.71 &1.1 & 6.2\\
$p_{\rm turb}$ & $10^6$ K~cm$^{-3}$ & 9.4 & 3.0 & 3.5 & 1.8 & 25.0 & 12.0 & 3.9 &1.8 &19.0\\
$\Delta v^2_{\rm turb}$ & km$^2$~s$^{-2}$ & 38.5 & 16.7 & 39.9 & 29.3 & 32.1 & 101.4 & 38.1 &14.2 &48.4\\
$N_{\rm H}$ (160~$\mu$m) & 10$^{22}$~cm$^{-2}$ & 4.5 &3.3 &5.6 &4.2 &4.5 & &13.0 & 13.0 & 4.6\\
$N_{\rm H}$ ($^{13}$CO) & 10$^{22}$~cm$^{-2}$ & 1.3 &0.84 &0.36 &0.78 &1.3 & &1.6 & 1.3 & 0.4\\
\enddata
\tablecomments{$n$ is hydrogen volume density and $G_{\rm 0}$ is far-UV flux determined from the PDR ToolBox. 
Errors from fits are in parenthesis. $G_{\rm 0}$ (FIR) is the far-UV flux determined from the Herschel data. $T_{\rm ex}$(\cii\,) is the \cii\ excitation temperature. $p_{\rm th}$ and $p_{\rm turb}$ are the thermal and turbulent pressures. $\Delta v^2_{\rm turb}$ is the turbulent velocity in km$^2$~s$^{-2}$. $N_{\rm H}$ (160~$\mu$m) and $N_{\rm H}$ ($^{13}$CO) are hydrogen column densities estimated using the 160~$\mu$m Herschel data and the $^{13}$CO APEX data. The columns p3, p3v1 and p3v2 correspond to the intensities integrated over the velocity window of -30 to 30~km~s$^{-1}$, -15 to 5~km~s$^{-1}$ and 5 to 20~km~s$^{-1}$, respectively.}
\label{tab:phys-val}
\end{deluxetable*}


We used two complementary fitting techniques from the PDR Toolbox to estimate the radiation field and H density, $G_{\rm 0}$ and $n$. The first is the Levenberg-Marquardt least squares fit (LSQ) method, which finds ``best-fit'' $G_{\rm 0}$ and $n$ by minimizing the $\chi^2$ of the observed intensity ratios versus the predicted model intensity ratios, weighted by the inverse square of the observational errors. 
The second method is a Markov Chain Monte Carlo (MCMC) method to determine the posterior probability density functions (PDFs) of $G_{\rm 0}$ and $n$. 
The MCMC results can been visualised through corner plots \citep{Foreman-Mackey2016}, which display the PDFs of $G_{\rm 0}$ and $n$ and the sampling distribution. As these distributions are approximately Gaussian, we characterise them on the overlay plots as 1 sigma contours in the estimated $G_{\rm 0}$ and $n$ values for each region (Figures \ref{fig:spaghetti-4-regs} and \ref{fig:spaghetti-h2}). As the estimated $n$ and $G_{\rm 0}$ from the two methods are consistent, we report the values derived by the MCMC method.

Although all intensity ratios enter into the calculation of the physical parameters, the ratios with the smallest error bar (the thinnest curve in overlay plots) dominate the fits. It can be seen in Figs.~\ref{fig:spaghetti-4-regs} and \ref{fig:spaghetti-h2} that the \cii\,/CO ratio plays the most important role in determining the physical conditions for different regions. Also, in certain cases, two points of intersection exist, but the \cii\,/CO dominates the determination of the physical conditions. The results are summarised in Table \ref{tab:phys-val}.

\subsection{Temperatures and pressures}
\label{sec:temp-press}

Using the derived $n$ and $G_{\rm 0}$ values for different regions, we determined the PDR surface temperatures ($T_{\rm surf}$ from the $T_{\rm surf}$ map in the PDR Toolbox), which are the model gas temperatures at $A_{\rm v}$ = 0.01 and are characteristic of the \cii\ and \oi\ emitting regions (Table~\ref{tab:phys-val}). 
We also estimated the \cii\ excitation temperatures ($T_{\rm ex}$) assuming an optical depth of $\sim$ 3, which should be considered an upper limit for the entire mapped region of RCW~49 \citep{Tiwari2021}. The calculated $T_{\rm ex}$ (\cii\,) are given in Table~\ref{tab:phys-val}. It can be seen that the $T_{\rm ex}$ (\cii\,) values are less than the $T_{\rm surf}$ values. 
This could be because \cii\ is actually optically thin towards most of the lines of sight and a derivation assuming it to be optically thick results in lower excitation temperatures.
Alternatively, a low excitation temperature may indicate a density below the critical density. 

We also calculated thermal ($p_{\rm th}$) and turbulent ($p_{\rm turb}$) pressures (Table~\ref{tab:phys-val}) in the different regions of RCW~49 using:

\begin{equation}
    p_{\rm th} = nT_{\rm surf} \hspace{1mm}({\rm K\, cm^{-3}}),
\end{equation}

and, 

\begin{equation}
    p_{\rm turb} = \mu mn \left[\frac{\Delta v^2_{\rm turb}}{8\ln (2)k}\right]  \hspace{1mm}({\rm K\, cm^{-3}}),
\end{equation}

where 
\begin{equation}
   \Delta v^2_{\rm turb} =  \Delta v^2_{\rm FWHM} - [8\ln (2)kT_{\rm surf}/m_{\rm c}]. 
\end{equation}

Here, $n$ is density 
(values from Table~\ref{tab:phys-val}), $T_{\rm surf}$ is surface temperature 
(values in Table~\ref{tab:phys-val}), $\mu$ = 1.3 is the mean molecular weight, $m$ is the hydrogen mass, 
 $\Delta v_{\rm FWHM}$ is full-width at half-maximum line width of \cii\ emission spectra (shown in Fig.~\ref{fig:spectra} and summarised in Table~\ref{tab:phys-val}), 
 $k$ is the Boltzmann constant 
and $m_{\rm c}$ is the carbon atom's mass. For determining $p_{\rm turb}$, we assumed $\Delta v^2_{\rm turb}$ to be dominated by turbulence.
However, the broad line widths of \cii\ could also be affected by velocity crowding along a line of sight. Thus, the $p_{\rm turb}$ estimated here should be considered as upper limits.  

We can see that the turbulent pressures are higher than the thermal pressures and thus dominate the dynamics of the gas in all regions of RCW~49. This is similar to the findings towards the stellar feedback driven shells of RCW~49 and RCW~36 (\citealt{Tiwari2021} and Bonne et al., submitted). When comparing the $p_{\rm th}$ and $p_{\rm turb}$ values derived for the shell in this work with that from \citet{Tiwari2021}, we find that the values estimated here are lower by up to a factor of $\sim$ 1.6. This difference is because in \citet{Tiwari2021}, we derived the physical parameters using the average intensity values of \cii\ and $^{12}$CO towards the entire shell of RCW~49, while in this work, we used the observed intensity values of \cii\,, \oi\,, $^{12}$CO and FIR towards a specific line-of-sight. Another reason is that in \citet{Tiwari2021}, the $G_{\rm 0}$ values towards the shell were estimated using the synthetic spectra for the stellar population of RCW~49 using the PoWR stellar atmosphere grids \citep{Sander2015}, while in this work we used the PDR ToolBox to derive $G_{\rm 0}$ values. Also, the PDR models we used in this work are newly updated (``wk2020'') compared to the ``wk2006'' models used in \citet{Tiwari2021}.

\subsection{Hydrogen nucleus column densities}
\label{sec:colden}

We estimated the dust column densities towards different regions of RCW~49 from the dust spectral energy distribution (see \citealt{Tiwari2021} for details).
The derived dust optical depths are converted to $N_{\rm H}$ (Table~\ref{tab:phys-val}) using the \citet{Draine2003} $R_{\rm v}$ = 3.1 value of the dust extinction cross section per hydrogen nucleus at 160~$\mu$m. We also calculated $N_{\rm H}$ using the $^{13}$CO data towards the same lines of sight. We first estimated the excitation temperature using the optically thick $^{12}$CO data, which can then be used to determine $^{13}$CO optical depths. Finally, using the $^{13}$CO integrated intensities along with the excitation temperatures and optical depths, we estimated the $^{13}$CO column densities, which can be converted to H$_{2}$ column densities (Table~\ref{tab:phys-val}), adopting a H$_2$/$^{13}$CO abundance of 6.1 $\times$ 10$^5$ \citep{Milam2005,Tielens2005}.  
The column densities derived from the dust are in general higher than the ones derived from $^{13}$CO.
This may indicate the presence of CO-dark gas along the line of sight. Alternatively, column density estimation from 70 and 160~$\mu$m data can include foreground and background contribution of up to 30\% \citep{Tiwari2021}.

\begin{figure*}[t]
\centering
\includegraphics[height=0.26\textwidth]{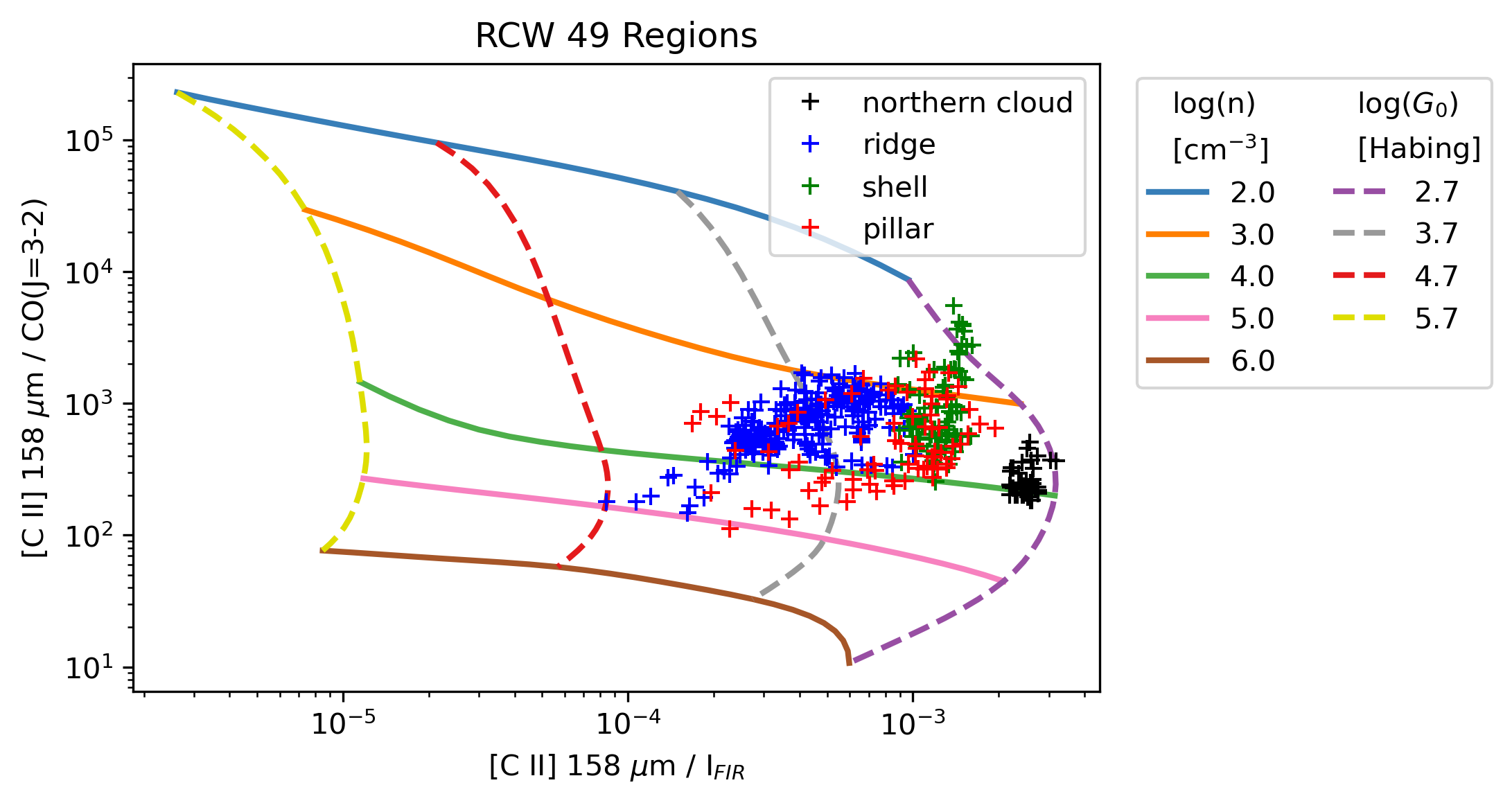}
\includegraphics[height=0.26\textwidth]{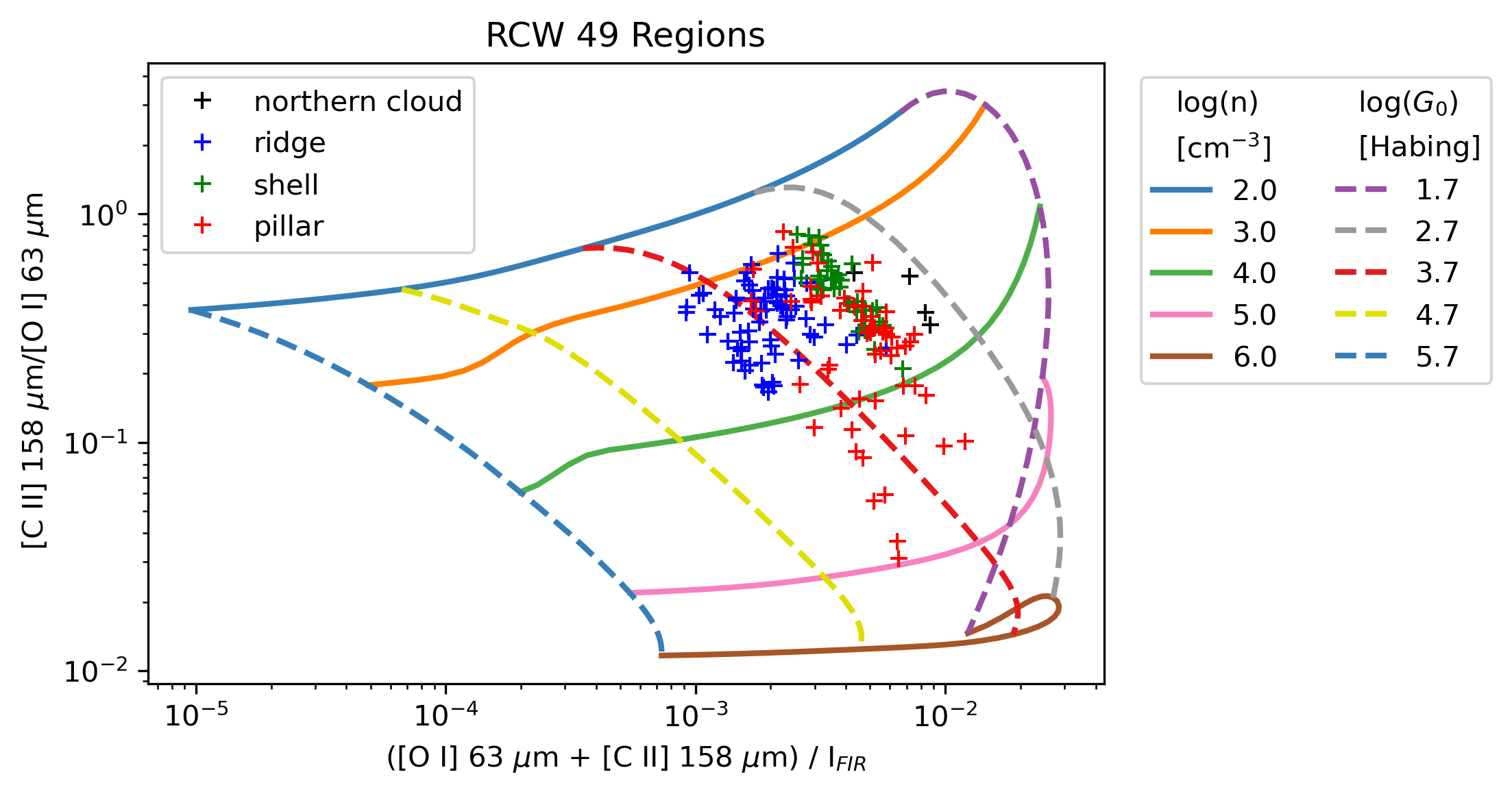}

\caption{The $n$ and $G_{\rm 0}$ model phase-space for \cii\,, \oi\,, $^{12}$CO and FIR intensities. left panel: modeled phase-space with respect to \cii\,/CO(3-2) vs. \cii\,/FIR. Right panel: modeled phase-space with respect to \cii\,/\oi\, vs. \oi\, + \cii\,/FIR. Data points are shown in colored (different regions) markers. The solid and dashed colored curves depict constant $n$ and $G_{\rm 0}$ values, respectively.     \label{fig:ph-sp-regs}}
\end{figure*}

\begin{figure*}[t]
\centering
\includegraphics[height=0.26\textwidth]{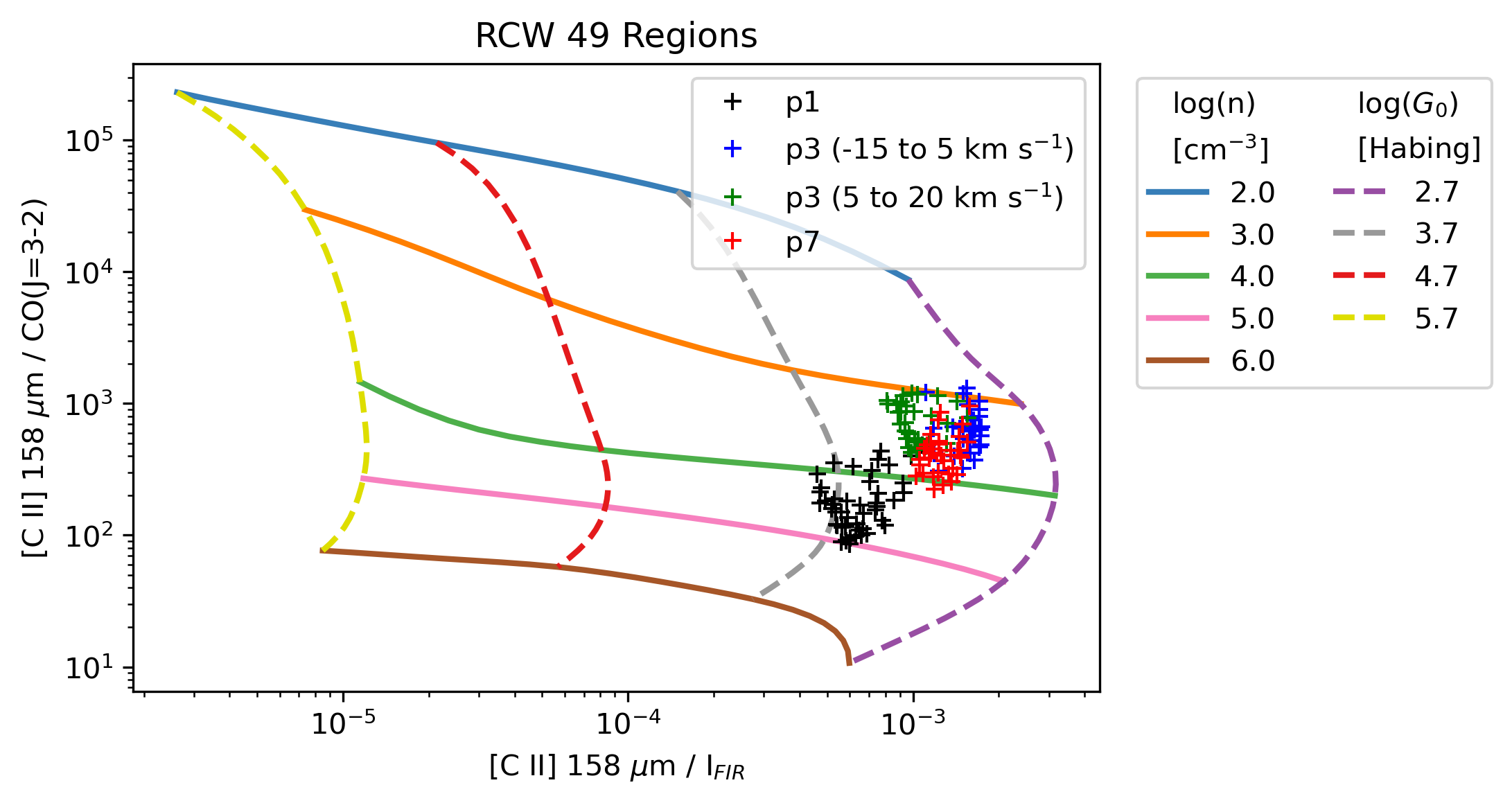}
\includegraphics[height=0.26\textwidth]{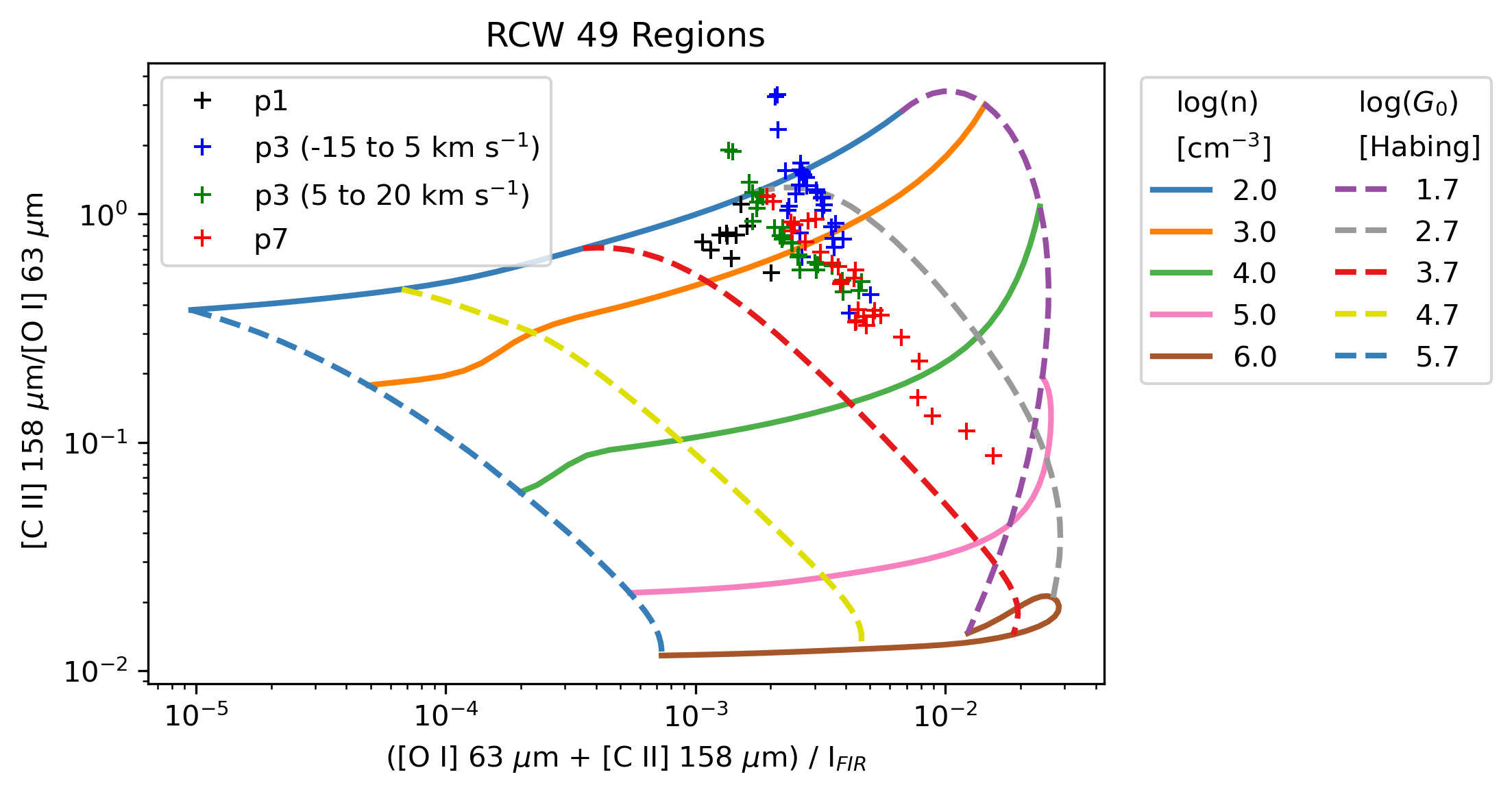}

\caption{The $n$ and $G_{\rm 0}$ model phase-space for \cii\,, \oi\,, $^{12}$CO and FIR intensities. left panel: modeled phase-space with respect to \cii\,/CO(3-2) vs. \cii\,/FIR. Right panel: modeled phase-space with respect to \cii\,/\oi\, vs. \oi\, + \cii\,/FIR. Data points are shown in colored (different regions) markers. The solid and dashed colored curves depict constant $n$ and $G_{\rm 0}$ values, respectively.     \label{fig:ph-sp-spit-pos}}
\end{figure*}

\subsection{Phase-space diagrams}\label{sec:phase-space-diags}

To examine the dispersion in physical properties across the mapped regions, we extracted all the data points from the \cii\,, CO, \oi\ and FIR intensity maps and overlaid their ratios on the $n$ and $G_{\rm 0}$ modeled phase-space (Figs.~\ref{fig:ph-sp-regs} and \ref{fig:ph-sp-spit-pos}). 
These specific diagnostic plots were selected for this analysis because the line ratios involved separate out $n$ and $G_0$ reasonably well into orthogonal directions. Specifically, the \cii\,/CO (3-2) and the \cii\,/FIR ratios are sensitive to $G_{\rm 0}$ while the \cii\,/\oi\, and (\oi\,+\cii\,)/FIR probe $n$. These diagnostic plots allow us to generalize the results of the detailed analysis of the specific points in Sects.~\ref{sec:overlay} to the whole region.

All the data were regridded to the same angular resolution of $\sim$ 20$\arcsec$ and filtered such that only data with values $> 3\sigma$ are included. For \cii\ and CO, most of the data points had values $> 3\sigma$. However, for \oi\,, the number of such data points decreased owing to lower sensitivity, thus, we used upper limits in certain cases.

The individual points in these regions cover a wide range in $n$ and $G_{\rm 0}$ (Table~\ref{tab:phys-val-ph-sp}).
The estimated range of $G_{\rm 0}$ values derived from ratios using \cii\,, $^{12}$CO and FIR intensities is similar to those derived using \cii\,, \oi\ and FIR intensities. However, the estimated $n$ values differ in certain cases. Specifically, in p1, the $n$ values in the left panel are higher by almost two magnitudes compared to the right panel. There are also significant differences in the ridge and p7. 
This is most likely because of the lower sensitivity of the \oi\ data towards these regions, resulting in a bias toward higher density regions.

\begin{deluxetable*}{l c c c c c c c c c}
\tablecaption{Physical conditions estimated by phase-space diagrams.}
\label{tab:phys-val-ph-sp}
\tablehead{
 \colhead{Parameter} & 
 \colhead{units} & 
 \colhead{northern cloud} & 
 \colhead{ridge} & 
 \colhead{shell} & 
 \colhead{pillar} & 
 \colhead{p1} &
 \colhead{p3v1} & 
 \colhead{p3v2} & 
 \colhead{p7}
}
\startdata
{$n$} (\cii\,-CO-FIR) & 10$^3$~cm$^{-3}$ & 5-10  & 1-100 & 0.5-5 & 1-100  &10-100  & 1-10&1-10  &1-10  \\
{$n$} (\cii\,-\oi\,-FIR) & 10$^3$~cm$^{-3}$ & 5-7 & 1-10  &  1-10 &  1-100 &0.1-1  & 0.5-5 & 0.5-5 & 0.1-70  \\
{$G_{\rm 0}$} (\cii\,-CO-FIR) & 10$^3$ Habing units & 0.5-0.7 & 1-50 & 0.5-1 & 1-10&0.8-6  &0.6-0.7 &1  & 0.7-1 \\
{$G_{\rm 0}$} (\cii\,-\oi\,-FIR) & 10$^3$ Habing units &  0.5-0.7  &  3-10&1 &0.7-10 & 1-4 &0.6-1 & 1 & 1 \\
\enddata
\tablecomments{Parameters with \cii\,-CO-FIR correspond to the values estimated from the left panels of Figs.~\ref{fig:ph-sp-regs} and \ref{fig:ph-sp-spit-pos}. Similarly, parameters with \cii\,-\oi\,-FIR correspond to the values estimated from the right panels of Figs.~\ref{fig:ph-sp-regs} and \ref{fig:ph-sp-spit-pos}. }
\label{tab:phys-val-ph-sp}
\end{deluxetable*}


\section{Discussion}\label{sec:discussion}

\subsection{Overlay plots vs. phase-space diagrams}\label{sec:overlay-pl-vs-ph-sp-pl}

In Sect.~\ref{sec:overlay}, we used the LSQ and MCMC method to find the best point of intersection of all the observed ratios in different positions of RCW~49, while in Sect.~\ref{sec:phase-space-diags}, we used phase-space diagrams on a modeled grid of different combinations of ratios of \cii\,, \oi\, $^{12}$CO and FIR intensities. 
Although the phase-space diagrams are a good way to visualize the entire data ($\sim$ 30--230 \cii\ and $^{12}$CO spectra and $\sim$ 3--70 \oi\ spectra for different regions) in $n$--$G_{\rm 0}$ space and also visualizes gradients in a specific region, the estimations are biased based on the combination of ratios used for specific species. For instance, the \cii\,/CO(3-2) ratio is more sensitive to $G_{\rm 0}$, while the \oi\,/\cii\ ratio is more sensitive to $n$. Another point to note is that observed \oi\ observations have lower sensitivity than the other tracers and thus, the phase-space probed by the \oi\,/\cii\ ratio will have larger errors compared to the phase-space probed by the \cii\,/CO(3-2) ratio.
In contrast, deriving the physical parameters using all the ratios (four in Fig.~\ref{fig:spaghetti-4-regs} and five in Fig.~\ref{fig:spaghetti-h2}) simultaneously towards a representative position in different regions (as done in Sect.~\ref{sec:overlay}) should be considered a more accurate approach for PDR analysis. 
Another way to compare the physical parameters derived from the overlay plots and the phase-space diagrams is by identifying the specific data points used in the overlay plots (from Fig.~\ref{fig:spaghetti-4-regs} and \ref{fig:spaghetti-h2}) in the phase-space diagrams. 
These figures are shown and discussed in detail in Appendix ~\ref{app:ph-sp-ov-dt-pts} and in Figs.~\ref{fig:ph-sp-regs-1val} and \ref{fig:ph-sp-spit-pos-1val}. We find that the data points used to do the analysis of the overlay plots lie on the extreme ends of the phase-space probed by the entire data sets of different regions, which might imply that the results obtained from the overlay plots are not characteristic of the entire regions. However, as mentioned in Sect.~\ref{sec:spectra}, these specific lines of sight were chosen for the overlay plots to ensure that clump emission is excluded, which can interfere with the PDR analysis and give misleading results. Thus, we believe that the overlay plots (results summarised in Table~\ref{tab:phys-val}) towards these representative lines of sight are a better way to estimate the physical parameters and we use them as a guide to understand stellar feedback in RCW~49.

\begin{figure}[t]
\centering

\includegraphics[width=85mm]{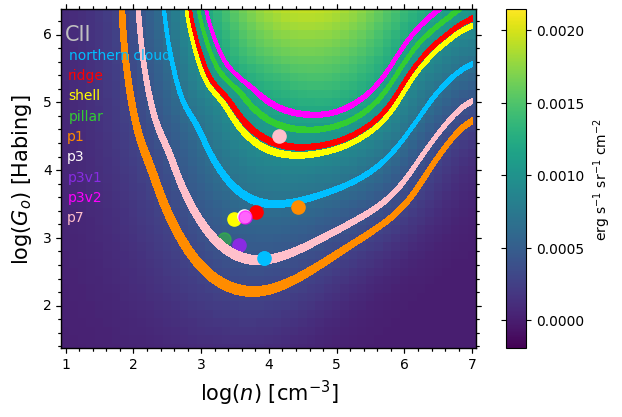}

\caption{\cii\ modeled intensity map as a function of $G_{\rm 0}$ and $n$. The map is highlighted with the observed intensity contours as well as marked with the model predicted intensity values for different regions in different colors. The thickness of the contours correspond to their observational error bars.    \label{fig:cii-obs-pred}}
\end{figure}

\subsection{Observed versus model intensities}
\label{sec:obs-vs-exp-int}

As described in Sect.~\ref{sec:overlay}, the PDR Toolbox derives $n$ and $G_{\rm 0}$ (given in Table~\ref{tab:phys-val}) from the ratios of \cii\,, \oi\,, CO(3-2), H$_2$ 0-0 S(1), H$_2$ 0-0 S(2) and FIR intensities. We will now compare the individual observed intensity values of different species with those predicted by the models.
For this, we extracted the intensity values of \cii\,, CO, \oi\,, H$_2$ 0-0 S(1) and H$_2$ 0-0 S(2) from the ``wk2020" PDR Toolbox models for the derived $n$ and $G_{\rm 0}$ values for different regions.  
We can see from the Table~\ref{tab:int-exp} that the model predicted intensity values of different species differ from their observed values. The differences are most significant for p3 and p7 positions. 

These differences can be visualised in Fig.~\ref{fig:cii-obs-pred}, where the \cii\ modeled intensity map is highlighted with the observed intensity contours along with the predicted \cii\ intensities for different regions in RCW~49. 
It is clear that the observational errors do not explain the difference between the predicted and observed \cii\ intensity values.

We ascribe these differences to the effects of geometry.
The current PDR models assume a face-on geometry and we can
check results for an edge-on geometry \citep[see][]{Pabst2017}. In this view,
the radiation is incident from the side, and the PDR layers are spread across
the sky with the atomic gas closest to the incident radiation and the molecular gas at greater distance. We calculate the line intensities from the (face-on) local emissivities integrated  along the line of sight while accounting for the line optical depths.  At a typical radiation field and density found in Table~\ref{tab:phys-val} ($G_{\rm 0}$ $\sim$ 3 $\times$ 10$^3$ Habing units and $n$ $\sim$  $\times$ 10$^3$~cm$^{-3}$), the \cii\ intensity is about a factor of 2 higher in the edge-on case compared to face-on due to the increased
column density along the line of sight. The \oi\ and $^{12}$CO lines are relatively unchanged since they are already optically thick and increasing the column density does little to increase the emitted line intensity. However, this would lead to disagreement with the observed intensity ratios.
Since the density is greater than the \cii\ critical density but less than the \oi\ and $^{12}$CO critical densities, an increase in the density by a factor of 2 does little to change \cii\ but increases \oi\ by a factor of $\sim$ 2 and $^{12}$CO by a factor of $\sim$ 2-3. Combining these two effects, the observed ratios can be explained by an edge-on model with a factor of $\sim$ 2 higher density.
Thus, the model fits to some of the regions could be improved and geometry introduces an uncertainty by a factor
of ${\sim} 2$ in density. Position p3 remains problematic with a
high observed \cii\ intensity. Such high intensities $> 10^{-3}$
erg ${\rm cm^{-2}}$ ${\rm s^{-1}}$ ${\rm sr^{-1}}$
typically indicate a large contribution from ionized gas
along the line of sight \citep{Seo2019}. 

Overall, we believe that geometrical effects are the main cause for the difference between the predicted and the observed intensity values for various species.

\begin{figure}[t]
\centering
\includegraphics[height=0.37\textwidth]{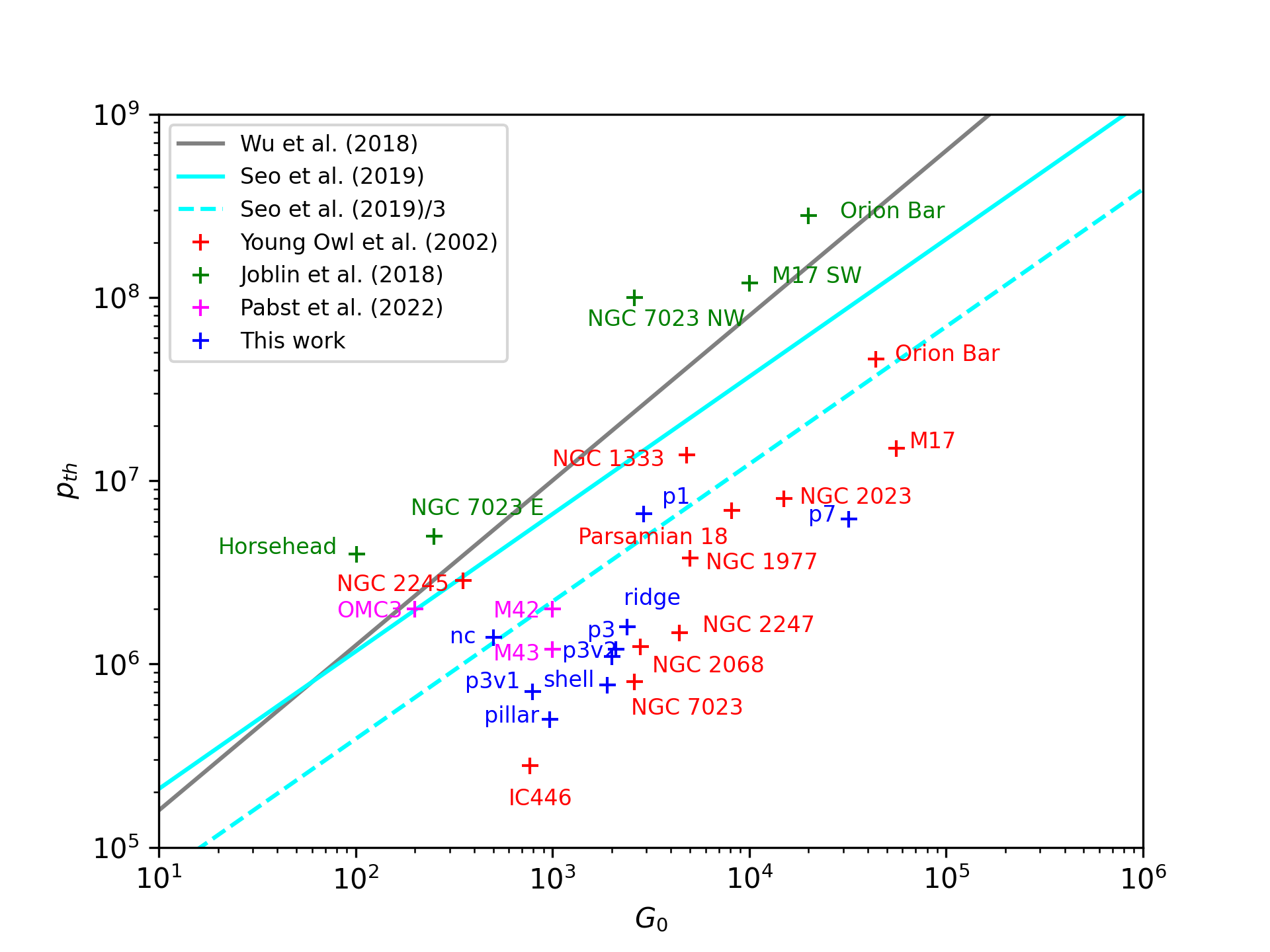}

\caption{$p_{\rm th}$ vs $G_{\rm 0}$ for different regions. The red, green, magenta and dark blue points correspond to regions studied by \citet{Young-Owl2002} (using FIR lines analysis), \citet{Joblin2018} (using high-$J$ $^{12}$CO analysis), \citet{Pabst2021} (using low-$J$ $^{12}$CO and \cii\ analysis) and this work, respectively. p3v1 and p3v2 correspond to the two velocity components of p3: -15 to 5 km~s$^{-1}$ and 5 to 20~km~s$^{-1}$, respectively. The black line refers to the fit by \citet{Wu2018} using mid to high-$J$ $^{12}$CO and \ci\ analysis. The light blue solid and dashed lines are for the analytic relation by \citet{Seo2019} and the relation divided by 3 to account for pressure equipartition, respectively.       \label{fig:pth-g0}}
\end{figure}

\subsection{Relationship between $p_{\rm th}$ and $G_{\rm 0}$}\label{sec:pth-vs-g0}
We studied the relationship between $p_{\rm th}$ and $G_{\rm 0}$ for different regions of RCW~49 (Table~\ref{tab:phys-val}) and for various PDRs studied in the literature.
Figure~\ref{fig:pth-g0} is adapted from \citet[Fig.~20]{Pabst2021}. We can see that all studies follow the general trend, where $p_{\rm th}$ increases with $G_{\rm 0}$. However, studies by \citet{Young-Owl2002} (red data points in Fig.~\ref{fig:pth-g0}) and \citet{Joblin2018} (green data points in Fig.~\ref{fig:pth-g0}) are somewhat shifted when compared to the regions of RCW~49 owing to higher $G_{\rm 0}$ and $p_{\rm th}$ values, respectively. \citet{Young-Owl2002} estimated higher values of $G_{\rm 0}$ for their sources compared to the values determined by \citet{Salgado2016}, \citet{Joblin2018} and \citet{Pabst2021}. \citet{Joblin2018} used high-$J$ $^{12}$CO lines for their analysis and therefore probe high density gas and/or clumps due to high critical densities of these lines. The relation between $p_{\rm th}$ and $G_{\rm 0}$ as derived by \citet{Wu2018} (black line in Fig.~\ref{fig:pth-g0}) also derived from high-$J$ $^{12}$CO lines analysis and is biased towards high density clumps. 

There is a simple relation (shown with solid light blue line in Fig.~\ref{fig:pth-g0}) between $p_{\rm th}$ and $G_{\rm 0}$ for a PDR in pressure equilibrium within a Stromgren sphere \citep{Young-Owl2002,Seo2019}.
The different regions of RCW~49 (dark blue data points in Fig.~\ref{fig:pth-g0}) lie close to the above analytical relation taking equipartition of thermal, turbulent, and magnetic pressures into account. This is similar to the sources studied by \citet{Pabst2021}. Thus, the physical conditions in different regions of RCW~49 are in agreement with basic Str{\"o}mgren relation for an \hii\ region.

\subsection{Impact of stellar feedback on different regions}\label{sec:feedback-on-regs}
In this section, we compare different regions of RCW~49 based on the derived physical conditions. The derived physical conditions in the seven regions of RCW~49 are indicative of different effects of stellar feedback on these regions. As justified earlier, we use the analysis reported in Sect.~\ref{sec:overlay} for the comparison.
We compare the regions p1, p3 and p7 separately from the northern cloud, ridge, shell and pillar because analysis of the former includes H$_2$ emission along a line of sight, while analysis of the latter excludes H$_2$ emission and includes emission only from specific velocity components corresponding to the given structure.

For $G_{\rm 0}$ values, the different regions can be placed in an increasing order: northern cloud with the lowest $G_{\rm 0}$, followed by the pillar, shell and the ridge with the highest $G_{\rm 0}$. Similarly, for the density, we can place the different regions in an increasing order:  pillar with the lowest $n$, followed by the shell and ridge and the northern cloud with the highest $n$. 

The northern cloud is located farthest from the Wd2 cluster, likely the reason that it has the lowest $G_{\rm 0}$ of the regions studied. The ridge, which is $\sim$ 40~pc away from the northern cloud \citep{Furukawa2009} is closest to the Wd2 cluster, corresponding to the highest $G_{\rm 0}$ value. Moreover, \citet{Furukawa2009} suggested that the northern cloud and the ridge collided $\sim$ 4~Myrs ago forming the Wd2 cluster (age $\sim$ 2--3~Myrs, \citealt{Furukawa2009,Tiwari2021}). The compression associated with this collision may be the cause for the higher densities derived for these regions. The high density of the ridge is also consistent with the ongoing star formation reported in it \citep{Whitney2004,Tiwari2021}. Also, the derived densities of the northern cloud and the ridge in this work are similar to the ones (3--8 $\times$ 10$^3$~cm$^{-3}$) derived by \citet{Ohama2010} using the LVG analysis of molecular emission lines.  
The mechanical feedback of the Wd2 cluster drives the expanding (at $\sim$ 13~km~s$^{-1}$) shell of RCW~49, which is $\sim$ 6~pc from the cluster. A secondary (younger and lower in mass) generation of star formation is taking place in the shell of RCW~49 (as discussed in \citealt{Whitney2004,Tiwari2021}). The pillar has similar $n$ and $G_{\rm 0}$ values (1.5 and 1.6 times lower) as the shell. From the pillar's morphology, it seems to be created by the Wd2 cluster. However, it also has the Wolf-Rayet star, WR20b, relatively closer to it and that could also be responsible for illuminating the pillar. Currently, we are unable to quantify the effects of radiative feedback from WR20b in defining the physical conditions of the pillar. But comparing the masses and bolometric luminosities of WR20b with that of Wd2 cluster, the stellar (radiative and mechanical) feedback from WR20b should be not more than 20\% of that from Wd2.

Among p1, p3 and p7, p7 has the highest $G_{\rm 0}$ value, while the -15 to 5~km~s$^{-1}$ velocity component of p3 has the lowest $G_{\rm 0}$ value. And p1 has the highest $n$ value and again the -15 to 5~km~s$^{-1}$ velocity component of p3 has the lowest $n$ value. P7 is the northernmost part of the shell of RCW~49 and in the west of it, the shell gets broken. We also located an early type O9V star close to this line of sight. Thus, the high $G_{\rm 0}$ value in p7 could be due to its proximity to this star. p1 is closest to the Wd2 cluster and has dense gas towards it (bright $^{12}$CO and 870~$\mu$m emission), suggesting a possibility of triggered star formation in its early stage. Towards position p3, we see a superposition of the shell (-15 to 5~km~s$^{-1}$) and the ridge (5 to 20~km~s$^{-1}$) components. The $G_{\rm 0}$ values for the shell component are lower than that of the ridge. This can be explained from their location with respect to the Wd2 cluster. As mentioned in the previous paragraph, the ridge is close to Wd2 while, the shell is about 6~pc away from it. The ridge component is also denser compared to the shell component in p3, which is consistent with its ongoing star formation.

\section{Conclusions}
\label{sec:conclusions}

We studied seven different regions of RCW~49 using the \cii\,, \oi\,, $^{12}$CO (3-2), H$_2$ 0-0 S(1) and H$_2$ 0-0 S(2) observations. Four of these regions: the northern cloud, ridge, shell and pillar have the \cii\,, \oi\, and $^{12}$CO (3-2) data towards them observed through the FEEDBACK program. These regions are both spatially and spectrally distinct. The other three regions: p1, p3 and p7 have the H$_2$ 0-0 S(1) and H$_2$ 0-0 S(2) data observed by the Spitzer telescope in addition to the FEEDBACK data. 
We presented the \oi\ and H$_2$ data (spectra and emission maps) towards RCW~49 for the first time in this paper. We found that the H$_2$ 0-0 S(1) emission distribution follows that of \cii\ while the H$_2$ 0-0 S(2) emission distribution follows that of CO, indicating that H$_2$ 0-0 S(2) arises from a denser gas compared to H$_2$ 0-0 S(1) in RCW~49. 

To determine the physical conditions in different regions of RCW~49, we compared our observations with the updated PDR models. We justified that the PDR analysis done using the overlay plots and data fits is a better technique than using phase-space diagrams to derive the physical conditions in the ISM. Based on the physical conditions, we studied the effects of stellar feedback on the evolution of different regions in RCW~49. We found that the ridge (closest to Wd2) has the highest $G_{\rm 0}$ value, while the northern cloud (farthest from Wd2) has the lowest $G_{\rm 0}$. However, both the northern cloud and the ridge have high densities, which is consistent with previous studies suggesting that these regions are part of large scale clouds, whose collision led to the formation of the Wd2 cluster. Also, there is evidence of ongoing star formation in the ridge. Among p1, p3 and p7, p1 has highest density and based on the bright $^{12}$CO and 870~$\mu$m emission towards it, we suggest early stage star-formation in this region. p7 has highest the FUV flux and this is attributed to the impinging radiation from an early type O9V star close to this position. 

We also estimated pressures towards these regions and found that the $p_{\rm turb}$ dominate (when compared to $p_{\rm th}$) the dynamics of the gas in RCW~49. Furthermore, the observed relationship between $p_{\rm th}$ and $G_{\rm 0}$ follow the theoretical relationship derived from the Str\"{o}mgren relation for an \hii\ region, which has also been observed toward several other PDRs.  \\

We would like to thank the referee for their constructive comments and helping to clarify the paper. 

This study was based on observations made with the NASA/DLR Stratospheric Observatory for Infrared Astronomy (SOFIA). SOFIA is jointly operated by the Universities Space Research Association Inc. (USRA), under NASA contract NNA17BF53C, and the Deutsches SOFIA Institut (DSI), under DLR contract 50 OK 0901 to the University of Stuttgart. upGREAT is a development by the MPI f\"ur Radioastronomie and the KOSMA/University of Cologne, in cooperation with the DLR Institut f\"{u}r Optische Sensorsysteme.

Financial support for the SOFIA Legacy Program, FEEDBACK, at the University of 
Maryland was provided by NASA through award SOF070077 issued by USRA.
M.W.P. acknowledges support for development of the PDR Toolbox from NASA ADAP \#80NSSC19K0573.

The FEEDBACK project is supported by the BMWI via DLR, Project Number
50 OR 1916 (FEEDBACK). N.S., R.S., and L.B. acknowledge support by the
Agence National de Recherche (ANR/France) and the Deutsche
Forschungsgemeinschaft (DFG/Germany) through the project ``GENESIS''
(ANR-16-CE92-0035-01/DFG1591/2-1).

This work was partly supported by the Collaborative Research Centre 956, 
funded by the DFG.

\bibliography{references1}{}
\bibliographystyle{aasjournal}

\appendix

\section{IR spectra}\label{app:ir-spec}

\begin{figure}[h]
\centering
\includegraphics[width=0.328\textwidth]{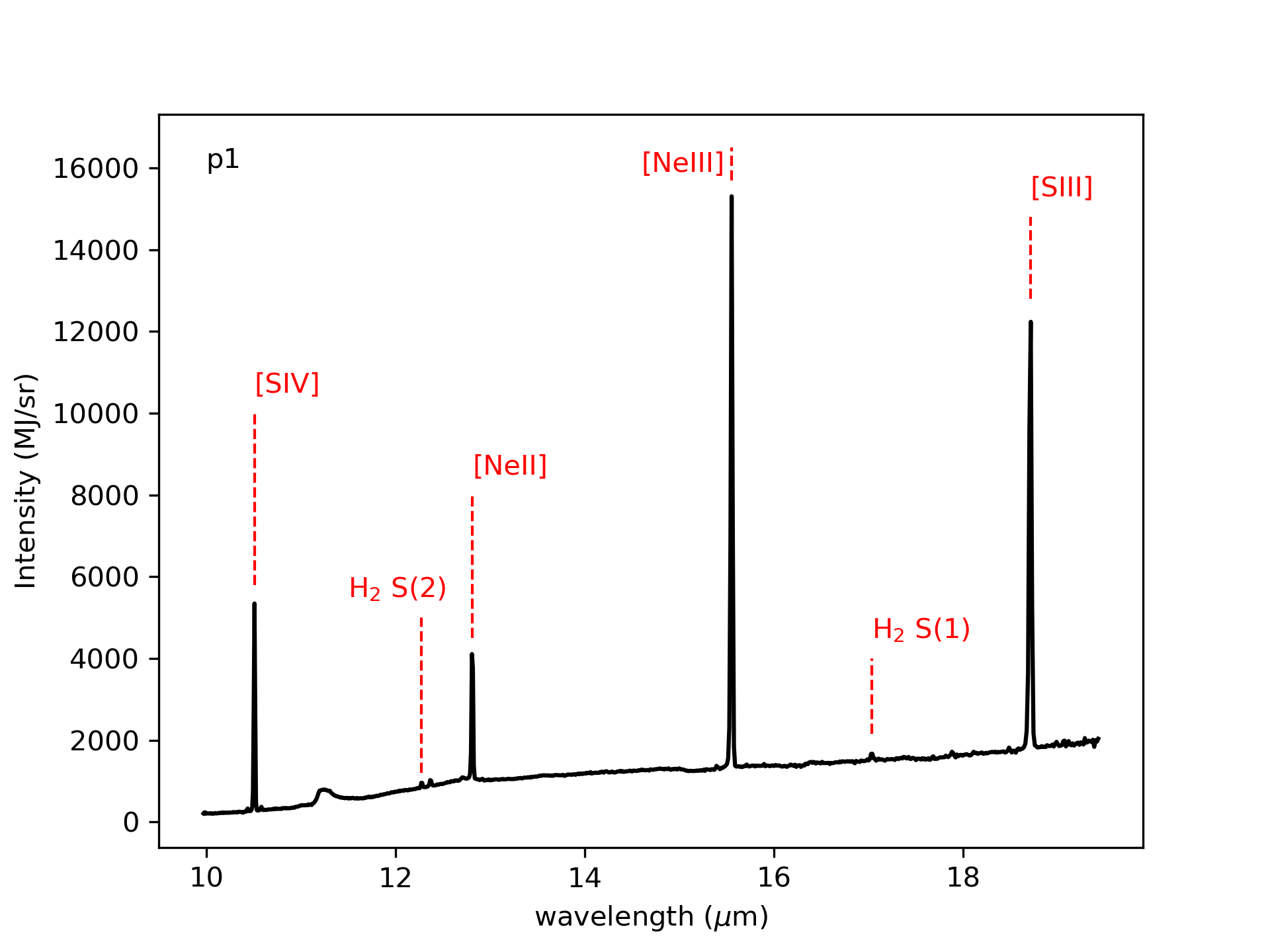}
\includegraphics[width=0.328\textwidth]{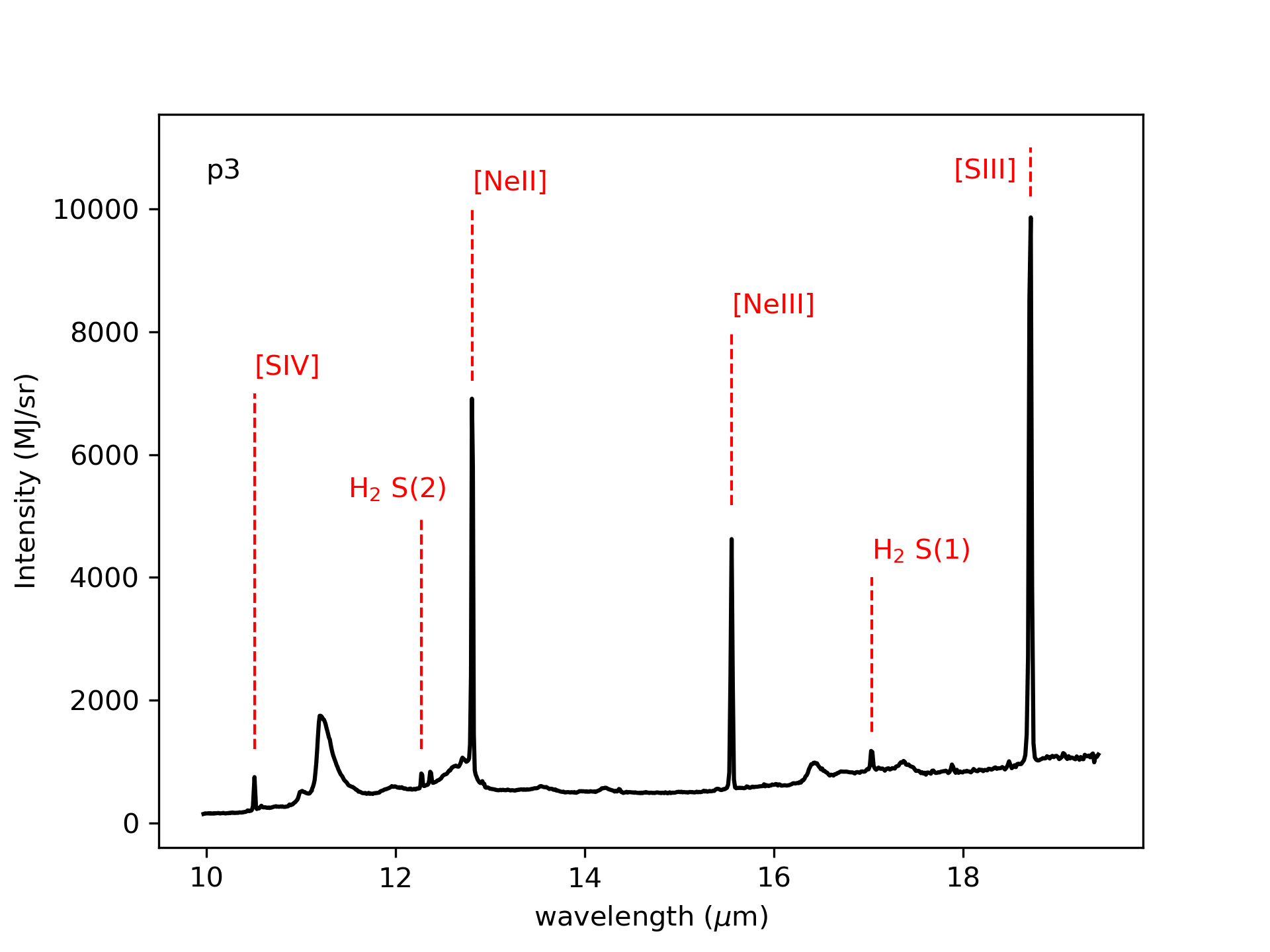}
\includegraphics[width=0.328\textwidth]{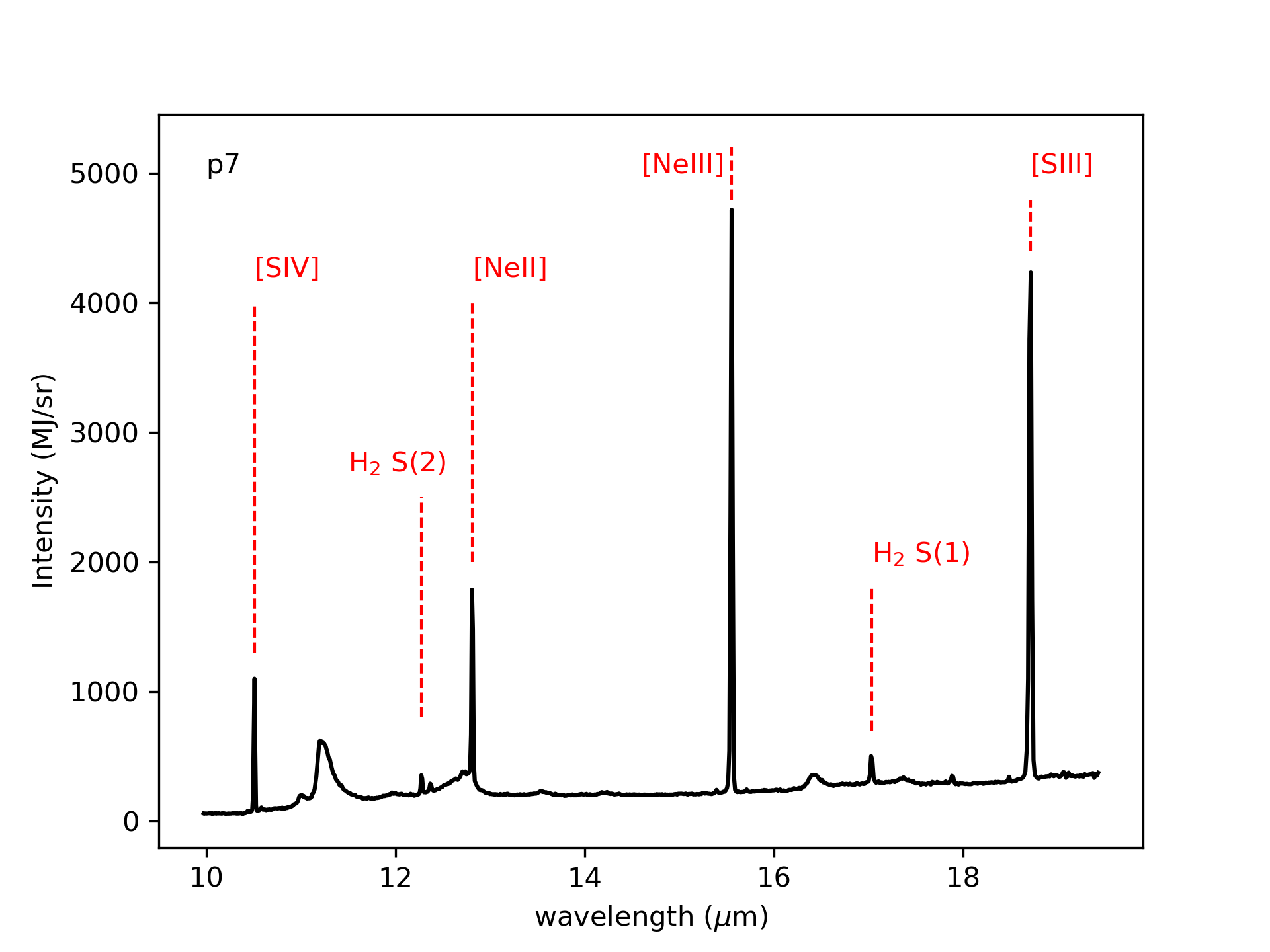}

\caption{Average IR spectra towards p1, p3 and p7 (regions shown in Fig.~\ref{fig:rgb}). The six species presented in this work are marked with red dashed lines.   \label{fig:ir-spec}}
\end{figure}

Figure~\ref{fig:ir-spec} displays the average spectra in $\sim$ 10--20~$\mu$m wavelength window towards p1, p3 and p7. H$_2$ lines are brightest in p3 by $\sim$ 1.5--3 times compared to p1 and p7. This is similar to our \cii\,, \oi\ and $^{12}$CO emission, which is brightest towards the ridge. The H$_2$ 0-0 S(1) line is similarly intense in p1 and p7, however the H$_2$ 0-0 S(2) line is $\sim$ 2 times brighter in p1 than in p7. This can be attributed to the fact that p1 is closer to Wd2 than p7 and is thus exposed to stronger radiation to heat the gas and excite the H$_2$ 0-0 S(2) line, which has an higher upper level energy.

For the ions, [\neii] and [\siii] are brightest in p3 and least bright in p7. This is similar to the H$_2$ lines. However, the relatively more ionised lines of [\neiii] and [\siv] are brightest in p1 and least bright in p7. This can again be explained because of the location of these regions with respect to Wd2: p1 is closest and p7 is farthest.

\begin{figure*}[h]
\centering
\includegraphics[height=0.26\textwidth]{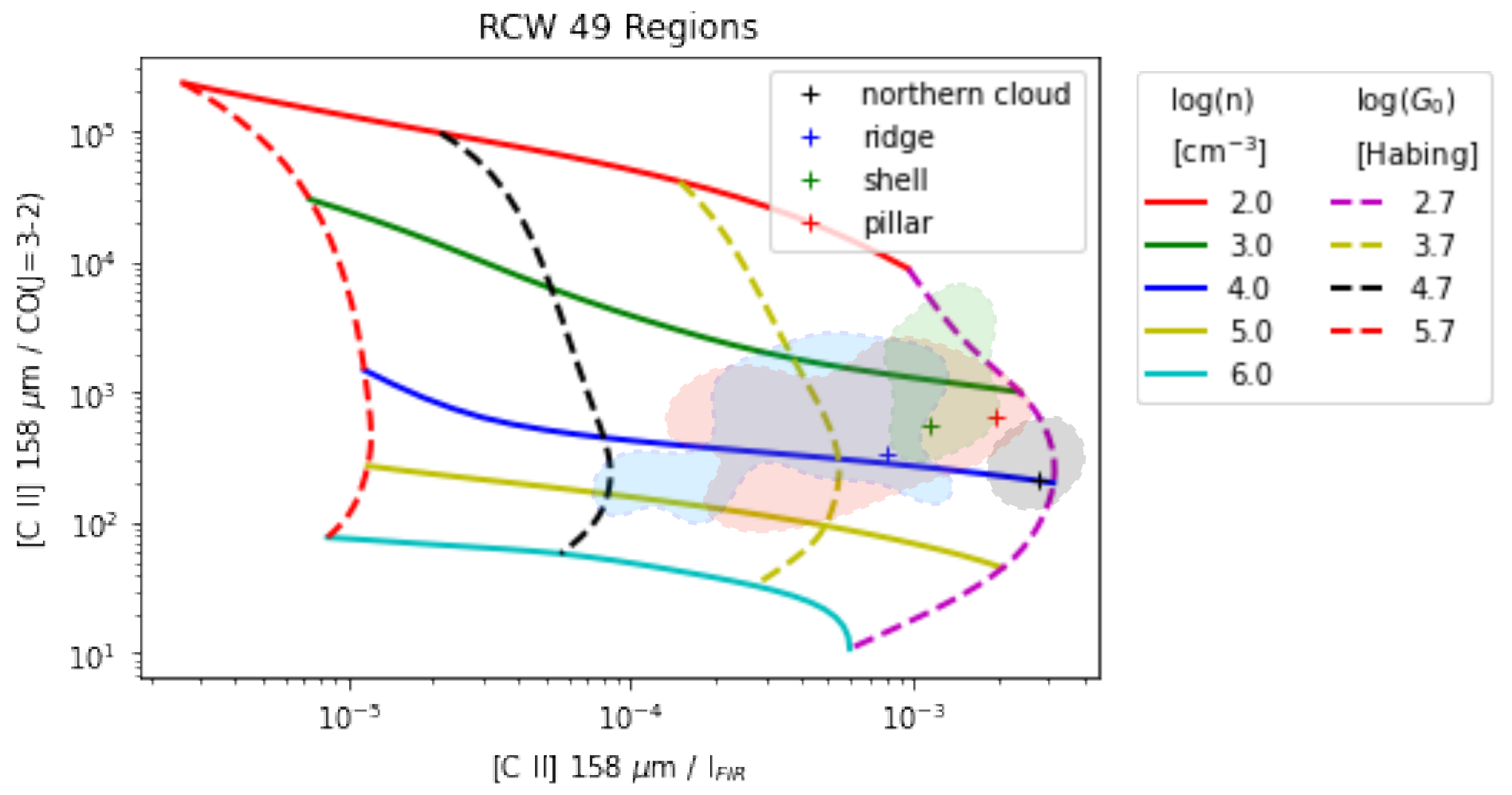}
\includegraphics[height=0.26\textwidth]{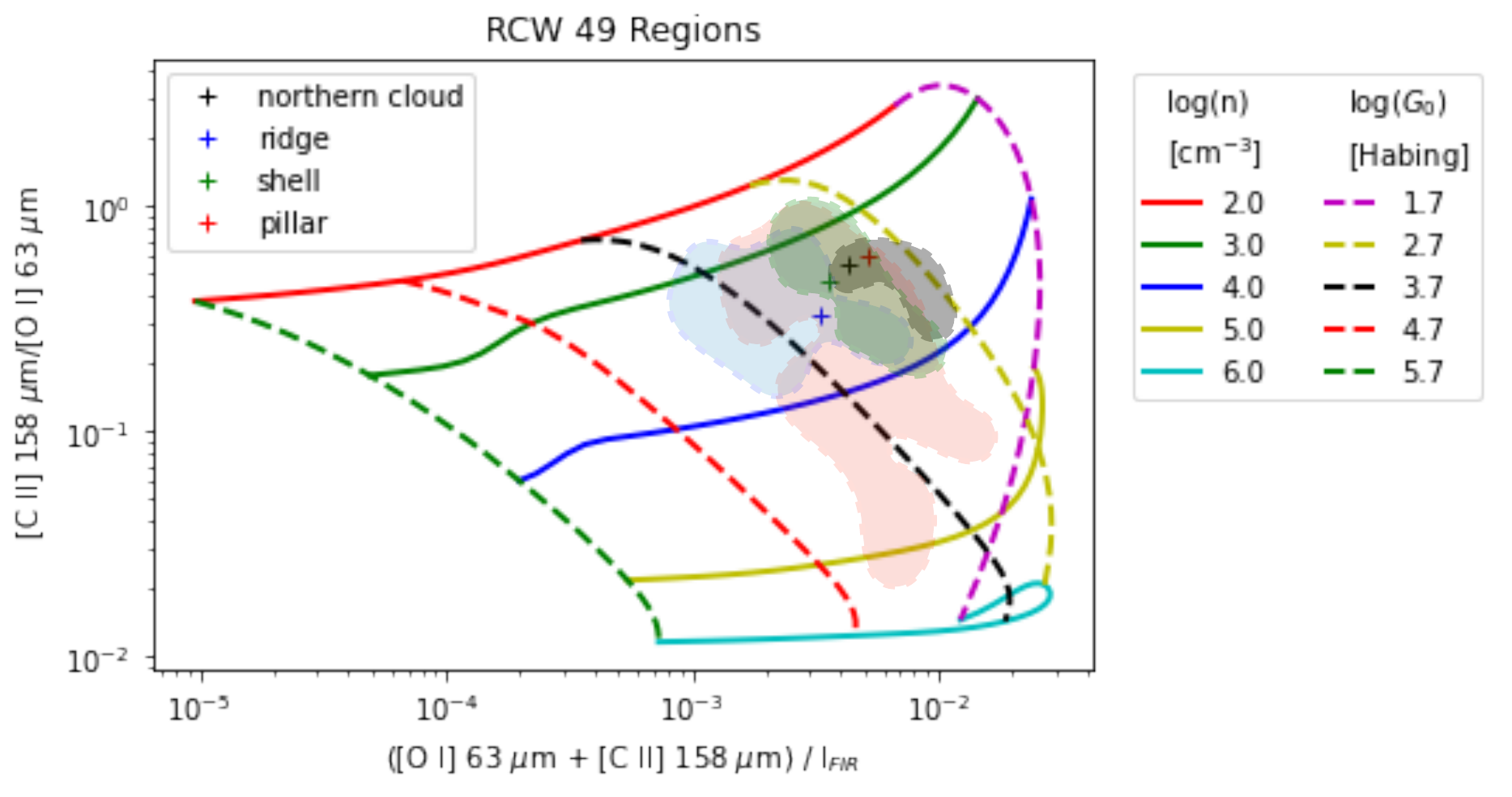}

\caption{The $n$ and $G_{\rm 0}$ model phase-space for \cii\,, \oi\,, $^{12}$CO and FIR intensities. left panel: modeled phase-space with respect to \cii\,/CO(3-2) vs. \cii\,/FIR. Right panel: modeled phase-space with respect to \cii\,/\oi\, vs. \oi\, + \cii\,/FIR. Data points are shown in colored (different regions) markers. The shaded regions correspond to the spread in the phase-space as probed by the entire dataset towards different regions as shown in Fig.~\ref{fig:ph-sp-regs}. The solid and dashed colored curves depict constant $n$ and $G_{\rm 0}$ values, respectively.     \label{fig:ph-sp-regs-1val}}
\end{figure*}

\begin{figure*}[h]
\centering
\includegraphics[height=0.26\textwidth]{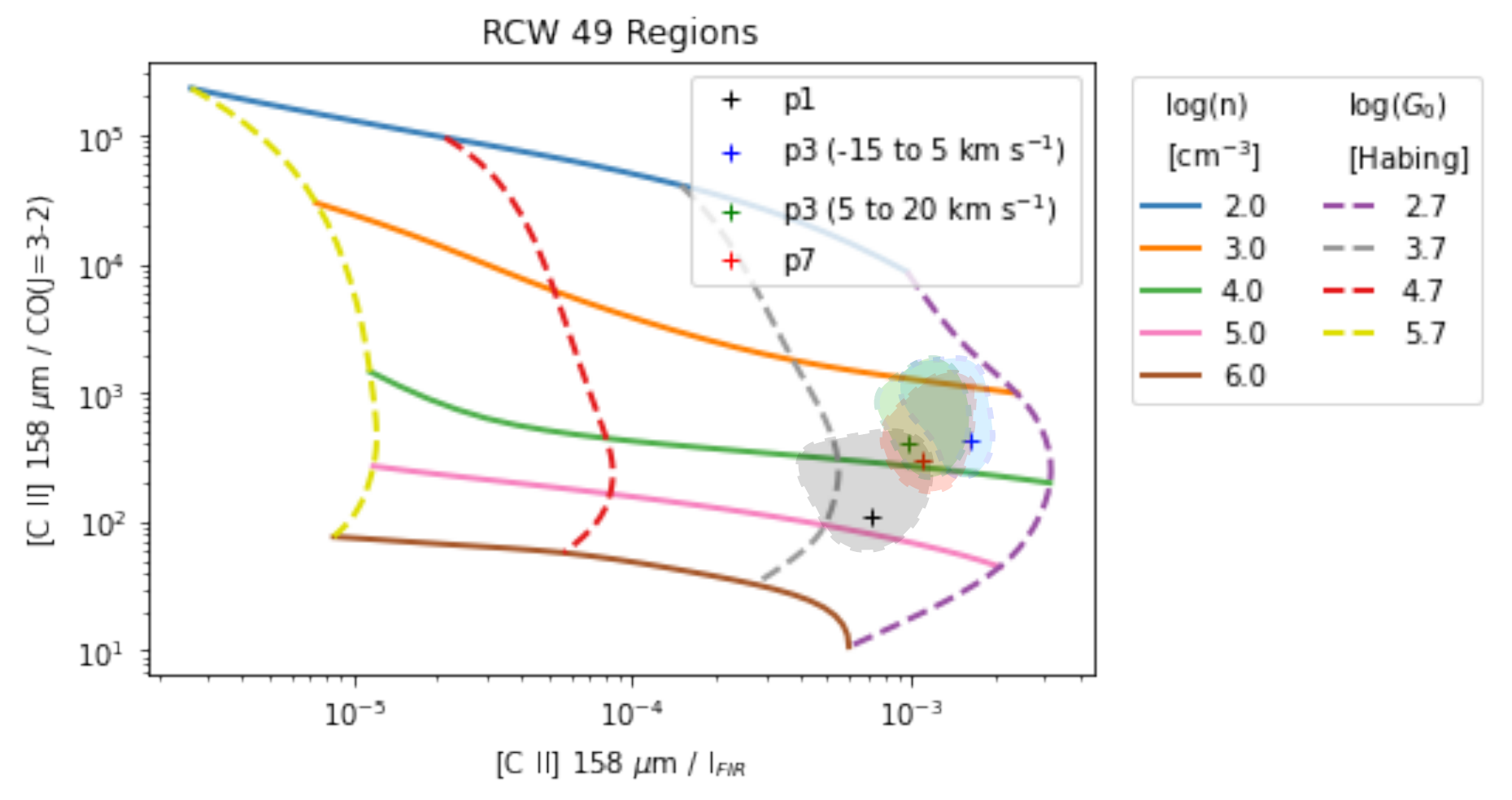}
\includegraphics[height=0.26\textwidth]{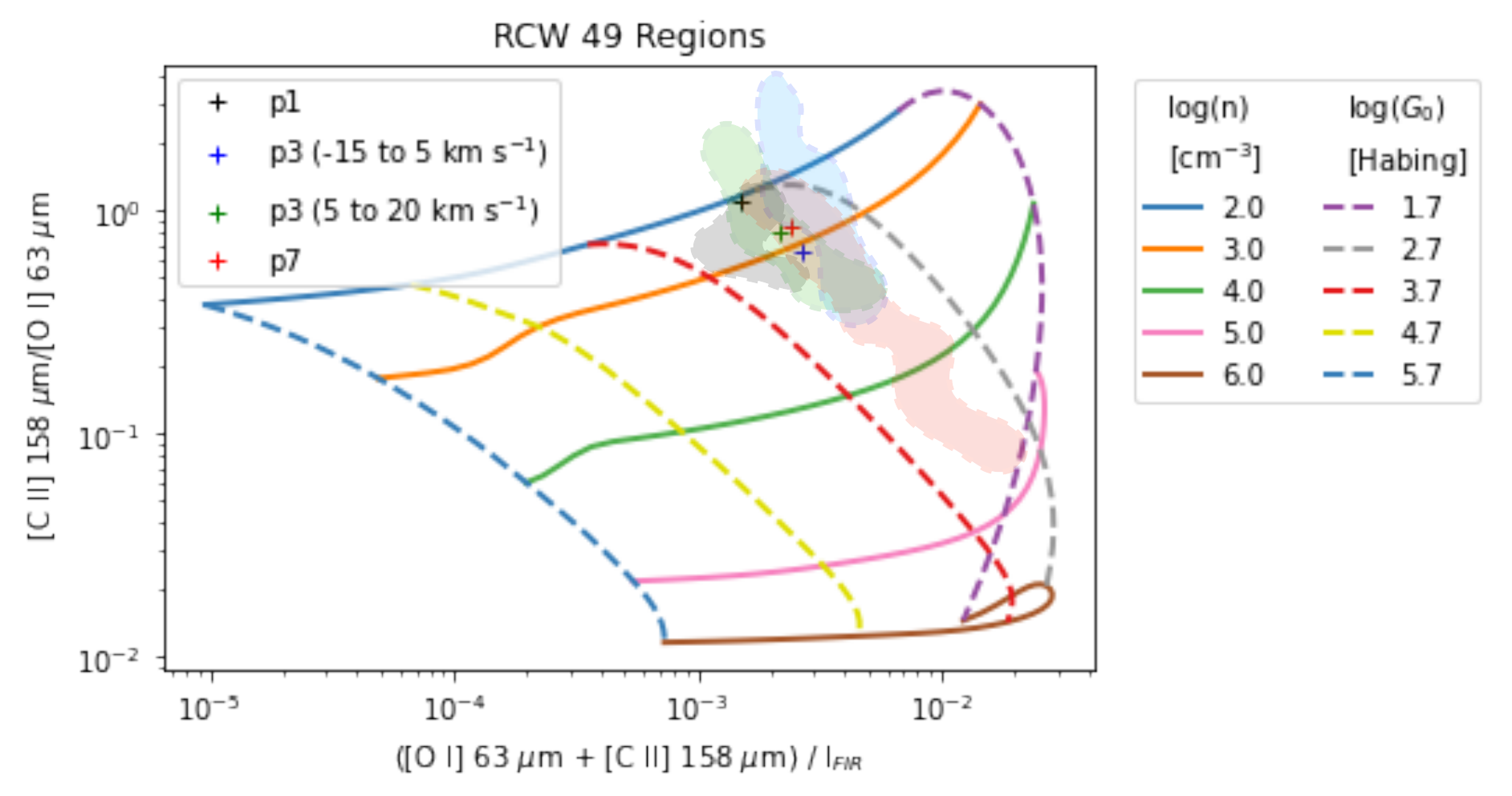}

\caption{The $n$ and $G_{\rm 0}$ model phase-space for \cii\,, \oi\,, $^{12}$CO and FIR intensities. left panel: modeled phase-space with respect to \cii\,/CO(3-2) vs. \cii\,/FIR. Right panel: modeled phase-space with respect to \cii\,/\oi\, vs. \oi\, + \cii\,/FIR. Data points are shown in colored (different regions) markers. The shaded regions correspond to the spread in the phase-space as probed by the entire dataset towards different regions as shown in Fig.~\ref{fig:ph-sp-spit-pos}. The solid and dashed colored curves depict constant $n$ and $G_{\rm 0}$ values, respectively.     \label{fig:ph-sp-spit-pos-1val}}
\end{figure*}

\section{Phase-space diagrams for specific data points}\label{app:ph-sp-ov-dt-pts}

Figures~\ref{fig:ph-sp-regs-1val} and \ref{fig:ph-sp-spit-pos-1val} show
phase-space diagrams overlaid with specific data points (colored crosses) selected for the PDR analysis presented in Sect~\ref{sec:overlay} and Figs.~\ref{fig:spaghetti-4-regs} and \ref{fig:spaghetti-h2}. These data points mostly lie close to the boundaries of the shaded areas (corresponding to the space probed by the entire data sets for different regions) in the phase-space diagrams. For instance, in the left panel of Fig.~\ref{fig:ph-sp-regs-1val}, the data points used to make the overlay plot for the ridge and pillar are probing lower values of $G_{\rm 0}$ compared to its shaded area, while for the shell, it is probing a higher value of $G_{\rm 0}$. Moreover, the data point corresponding to the pillar in the right panel of Fig.~\ref{fig:ph-sp-regs-1val} is probing lower value of $n$ compared to its shaded area. Another clear example can be seen in the right panel of Fig.~\ref{fig:ph-sp-spit-pos-1val}, where the data point corresponding to p7 probes lower density compared to its shaded area.

\end{document}